\documentclass{aa}  
\usepackage{graphicx}
\usepackage{txfonts}
\def\arcsec{$^{\prime\prime}$}
\def\arcmin{$^{\prime}$}
\def\degrees{$^{\circ}$}

\def\srm{$\sigma_{\rm RM}$}
\def\rmm{$\langle{\rm RM}\rangle$}
\def\absrmm{$\arrowvert \langle {\rm RM} \rangle  \arrowvert$}
\def\bm{$\langle\mathbf B\rangle$}
\begin{document}
   \title{The intracluster magnetic field power spectrum in Abell 2255}
   \subtitle{}

   \author{F. Govoni\inst{1}
          \and
          M. Murgia\inst{1,2}
          \and
          L. Feretti\inst{2}
          \and
          G. Giovannini\inst{2,3}
          \and
          K. Dolag\inst{4}
          \and
          G. B. Taylor\inst{5,6}
          }

   \offprints{F. Govoni, email fgovoni@ca.astro.it}

   \institute{INAF - Osservatorio Astronomico di Cagliari,
              Loc. Poggio dei Pini, Strada 54, I--09012 Capoterra (CA), Italy
           \and
              INAF - Istituto di Radioastronomia, 
              Via Gobetti 101, I--40129 Bologna, Italy
           \and     
              Dipartimento di Astronomia, 
              Univ. Bologna, Via Ranzani 1, I--40127 Bologna, Italy
           \and
              Max-Planck-Institut for Astrophysik, Garching, Germany
           \and
              Department of Physics and Astronomy, University of New Mexico, 
Albuquerque, NM 87131, USA  
           \and
              National Radio Astronomy Observatory, Socorro, NM 87801, USA   
              }

   \date{Received; accepted}

  \abstract
  {}
  {The goal of this work is to constrain the strength and structure
   of the magnetic field in the nearby cluster of galaxies A2255.
   At radio wavelengths A2255 is characterized by the presence of
   a polarized radio halo at the cluster center, a relic source at the cluster
   periphery, and several embedded radio galaxies. 
   The polarized radio emission from all these sources
   is modified by Faraday rotation as it traverses the magnetized intra-cluster 
   medium.  The distribution of Faraday rotation can be used to probe the magnetic 
   field strength and topology in the cluster. 
   }
  {For this purpose, we performed Very Large Array observations at 3.6 and 6\,cm 
   of four polarized radio galaxies embedded in A2255, obtaining detailed rotation
   measure images for three of them. We analyzed these data together with the very 
   deep radio halo image recently obtained by us.
   We simulated random 3-dimensional magnetic field models characterized by different 
   power spectra and produced synthetic 
   rotation measure and radio halo images. By comparing the simulations with the data
   we are able to determine the strength and the power spectrum of the intra-cluster 
   magnetic field fluctuations which best reproduce the observations.}
  {The data require a steepening of the power spectrum spectral index
   from $n=2$, at the cluster center, up to $n=4$,  at the cluster
   periphery and the presence of filamentary structures on large scales.
   The average magnetic field strength at the cluster center is
   2.5\,$\mu G$. The field strength 
   declines from the cluster center outward with an average magnetic field 
   strength calculated over 1 Mpc$^3$ of $\sim$1.2\,$\mu$G. 
  } 
   {}


   \maketitle
%

\section{Introduction}

The existence of magnetic fields associated with the intracluster
medium in clusters of galaxies is now well established through
different methods of analysis
(see e.g. the review by Govoni \& Feretti 2004, 
and references therein).
The strongest evidence for the presence of cluster magnetic fields
comes from radio observations.
Magnetic fields are revealed through the synchrotron
emission of cluster-wide diffuse sources, 
and from studies of the rotation measure (RM) of polarized radio
galaxies.
Other techniques, not discussed in this paper, include the study of
inverse Compton hard X-ray emission 
(e.g. Fusco-Femiano et al. 2004, Rephaeli et al. 2006), 
cold fronts  (e.g. Vikhlinin \& Markevitch 2002)
and magneto hydrodynamic simulations (e.g. Dolag et al. 2002).
 
Direct evidence for the presence of relativistic electrons and magnetic 
fields in clusters of galaxies comes from the detection, 
in an ever increasing number of 
galaxy clusters, of large-scale, diffuse, steep-spectrum synchrotron 
sources known as 'radio halos'  
or 'relics' depending on their morphology and location
(e.g. Giovannini \& Feretti 2002).
Under the assumption that radio sources are in a minimum energy condition,
it is possible to derive a zero--order estimate of the magnetic field
strength averaged over the entire source volume.
Typical equipartition magnetic fields, estimated
in clusters with wide diffuse synchrotron emission, are 0.1-1 $\mu$G.
However, the equipartition estimate is critically dependent on the low
energy cut-off of the relativistic electrons for steep spectrum 
sources such as radio halos. Since this quantity is not known
the equipartition estimates of the magnetic field strength in these
radio sources should be used with caution. 
 
The presence of a magnetized
plasma between an observer and a radio source 
changes the properties of the polarized emission from the
radio source.
Therefore a complementary set of information on clusters 
magnetic fields along the line-of-sight can be determined, in conjunction
with X-ray observations of the hot gas, through the analysis of the
RM of radio sources.
Many high quality RM images of extended radio galaxies
are now available in the literature (see e.g. the review by Carilli \& Taylor 2002, and 
references therein). These data 
are consistent with magnetic fields of a few $\mu$G 
throughout the clusters.
In addition, stronger fields exist in the inner regions of 
strong cooling core clusters.
In a few cases it has been possible to study the cluster magnetic field
in more detail by sampling several
extended radio galaxies located in the same cluster of galaxies
(e.g. Feretti et al. 1999, Taylor et al. 2001, Govoni et al. 2001a).

It is worth noting that the RM observed toward radio galaxies may not
be entirely representative of the cluster magnetic field
if the RM is locally enhanced by compression of the intracluster medium 
due to relative motions.
There are, however, several statistical arguments
against this interpretation.
In particular the statistical RM investigation 
of point sources (Clarke et al. 2001, Clarke 2004) shows a clear
broadening of the RM distribution toward small
projected distances from the cluster center indicating that most of the RM contribution comes
from the intracluster medium.  This study included background
sources, which showed similar enhancements as the embedded sources.

Recent work (e.g. En{\ss}lin \& Vogt 2003, Murgia et al. 2004) showed that detailed
RM images of radio galaxies can be used to infer 
not only the cluster magnetic field strength, 
but also the cluster magnetic field power spectrum.
Moreover, Murgia et al. (2004) pointed out that
morphology and polarization information of radio halos may
provide important constraints on the 
power spectrum of the magnetic field fluctuations 
on large scales. In particular, their simulations showed that if the
intracluster magnetic field fluctuates up to scales of some hundred
kpc, then steep magnetic field power spectra 
may give rise to detectable polarized emission in radio halos.

A2255 is a nearby (z=0.0806, Struble \& Rood 1999), 
rich cluster with signs of 
undergoing a merger event (e.g. Yuan et al. 2005,
Sakelliou \& Ponman 2006).
It is a suitable target to study the intracluster magnetic field
because it is characterized by the presence of
a diffuse radio halo source at the cluster center,
a relic source at the cluster periphery, and several 
embedded radio galaxies (Jaffe \& Rudnick 1979,
Harris et al. 1980, Burns et al. 1995, Feretti et al. 1997).
Govoni et al. (2005) found, for the first time
in a cluster, that the radio halo of A2255 shows 
filaments of strong polarized emission ($\simeq$ 20-40\%).
The distribution of the polarization angles in the filaments 
indicates that the cluster magnetic field 
fluctuates up to scales of $\sim$400 kpc in size.
In the rest of the cluster they did not detect significant diffuse polarized emission 
except in the brighter regions of the relic ($\simeq $15-30\%). 

Here we present multi frequency (3.6 and 6 cm)
Very Large Array (VLA\footnote{The Very Large Array is a facility of the National Science Foundation, operated under 
cooperative agreement by Associated Universities, Inc.}) observations of four polarized radio galaxies 
in the A2255 cluster.
We apply the numerical approach proposed by Murgia et al. (2004),
of analyzing the polarization properties of
both radio galaxies and halo, to
investigate the cluster magnetic field strength
and power spectrum.

The paper is organized as follows.
In Sect. 2 we discuss details about the radio observations and the data
reductions.
In Sect. 3 we present the total intensity and polarization properties
of the four radio galaxies at 3.6 and 6 cm.
In Sect. 4 we present the RM images, discuss 
the results, and consider the presence of a cluster magnetic field.
In Sect. 5, by following the same approach 
presented in Murgia et al. (2004), we introduce the
3D multi-scale magnetic field modeling used to determine
the intra-cluster magnetic field strength and structure.
in Sects. 6 and 7 we show the results obtained with a constant and
variable magnetic field power spectrum slope, respectively. 
In Sect. 8, we draw conclusions from this study.  

Throughout this paper we assume a $\Lambda$CDM cosmology with
$H_0$ = 71 km s$^{-1}$Mpc$^{-1}$,
$\Omega_m$ = 0.3, and $\Omega_{\Lambda}$ = 0.7.
At the distance of A2255, 1\arcsec corresponds to 1.5 kpc.

\section{Radio observations and data reduction}

\begin{table*}
\caption{Pointed VLA observations of radio galaxies in the A2255 cluster of galaxies.}          
\label{data}      
\centering          
\begin{tabular}{c c c c c c c c}     
\hline\hline       
Source       & RA           &  DEC    &  $\lambda$ & Bandwidth & Config. & Date & Duration     \\
             & (J2000)      & (J2000) &  (cm)      & (MHz)     &         &      & (Hours)  \\\hline                    
J1712.4+6401 & 17 12 24.6   & +64 02 08 & 3.6 & 50 & B,C  & Nov 99, Mar 00  & 1.1, 1.4  \\ 
             &              &           & 6   & 50 & B,C  & Nov 99, Mar 00  & 1.1, 1.4  \\
J1713.3+6347 & 17 13 16.5   & +63 47 42 & 3.6 & 50 & B,C  & Nov 99, Mar 00  & 0.9, 1.3   \\
             &              &           & 6   & 50 & B,C  & Nov 99, Mar 00  & 0.9, 1.3   \\
J1713.5+6402 & 17 13 29.3   & +64 02 49 & 3.6 & 50 & B,C  & Nov 99, Mar 00  & 0.7, 1.0   \\
             &              &           & 6   & 50 & B,C  & Nov 99, Mar 00  & 0.7, 1.0   \\
J1715.1+6402 & 17 15 09.0   & +64 02 54 & 3.6 & 50 & B,C  & Nov 99, Mar 00  & 1.0, 1.2   \\
             &              &           & 6   & 50 & B,C  & Nov 99, Mar 00  & 1.0, 1.2   \\
\hline                  
\multicolumn{7}{l}{\scriptsize Col. 1: Source; Col. 2, Col. 3: Pointing position (RA, DEC);
Col. 4: Observing wavelengths;}\\

\multicolumn{7}{l}{\scriptsize Col 5: Observing bandwidth; Col. 6: VLA configuration; 
Col. 7: Dates of observation; Col. 8: Time on source.}\\ 
\end{tabular}
\end{table*}

\begin{figure*}
\centering
\includegraphics[width=18cm]{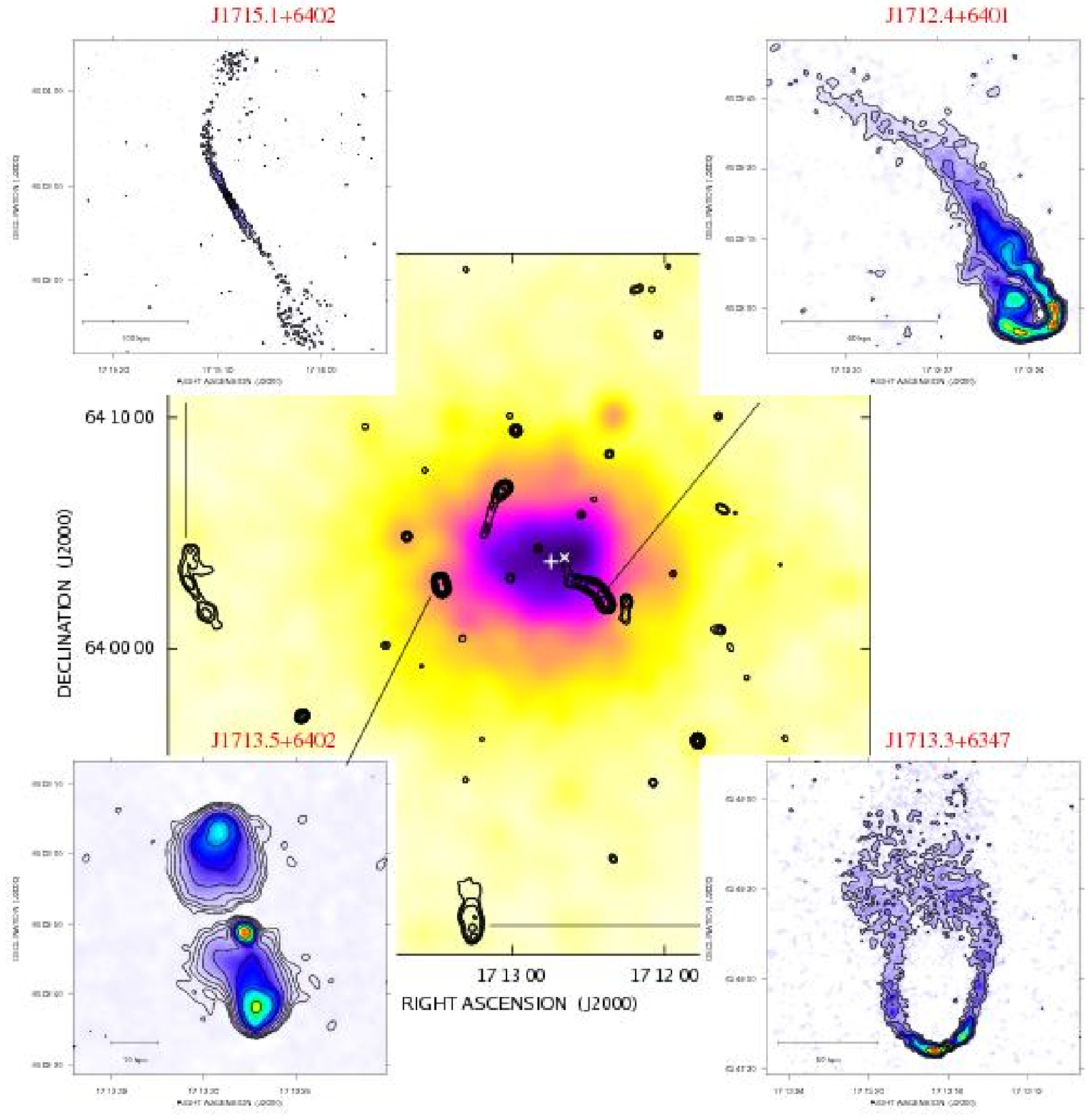}
\caption{Radio contours of the A2255 cluster of galaxies obtained at 20 cm 
(Govoni et al. 2005) overlaid on the ROSAT X-ray image.
The white symbols $+$ and $\times$ indicate the position of
the cluster X-ray centroid and peak respectively.
The radio image at 20 cm has a FWHM of $15''\times15''$.
The radio contours are:
0.4, 0.8, 1.6, 3.2, 6.4, 12.8, 25.6, and 51.2 mJy/beam.
The sensitivity (1 $\sigma$) is 0.016 mJy/beam.
For graphical reasons the first contour is at 25$\sigma$ therefore we
see only the radio galaxies 
of the cluster while the low brightness, diffuse emissions 
(radio halo and relic) are not visible here.
Pointed high resolution observations at 6 cm of four
cluster radio galaxies are inset.
The radio contours at 6 cm are:
0.06, 0.12, 0.24, 0.48, 0.96, 1.92, 3.84, and 7.68 mJy/beam.
See Table \ref{mappeIhigh} for more information (e.g. resolution, sensitivity, peak brightness)
of these observations.}
\label{6cm}
\end{figure*}

The four radio galaxies J1712.4+6401, J1713.3+6347, J1713.5+6402 and J1715.1+6402~ 
have been investigated with multi-frequency, high-resolution, VLA observations.
These sources have been selected on the basis of their high flux density ($>$80 mJy at 20 cm), 
extension, and the presence of polarized emission in the NVSS (Condon et al. 1998).
The details of the observations are provided in Table \ref{data}.
The sources were observed at two frequencies (4535/4885 MHz)
within the 6 cm band and at two frequencies (8085/8465 MHz) 
within the 3.6 cm band, both in the B and C configurations.
All observations were made with a bandwidth of 50 MHz.

The ($u,v$) data at the same frequencies
but from different configurations were first handled separately and then
combined.
The source $1331+305$ (3C286) was used as the primary 
flux density calibrator, and as an absolute reference for the 
electric vector polarization angle.
The phase calibrator was the
nearby point source $1642+689$, observed at intervals of about 30 minutes.
Calibration and imaging were performed with the
NRAO  
Astronomical Image Processing System
(AIPS), following standard procedures.
Self-calibration was applied to remove residual 
phase variations.

Total intensity images $I$ have been produced by averaging the two
frequencies in the same band 
while $U$ and $Q$ images have been obtained for each frequency separately. 
Images of polarized intensity $P=(Q^2+U^2)^{1/2}$, 
fractional polarization $FPOL=P/I$ and position angle of polarization
$\Psi=0.5\tan^{-1}(U/Q)$ were derived from the $I$, $Q$ and $U$ images.

\section{Total intensity and polarization properties}

Fig.~\ref{6cm} shows the contour image of A2255 (Govoni et al. 2005) 
at 20 cm overlaid onto the ROSAT PSPC image of the cluster.
The X-ray image is in the band 0.5-2 keV and has been
obtained from the ROSAT public archive
by binning the photon event table in pixels of 15$''$ 
and by smoothing the image with a Gaussian of $\sigma=30''$.
The centroid of the X-ray emission is approximately at
RA(J2000)=$17^h12^m45^s$, DEC(J2000)=64\degrees03\arcmin54\arcsec 
(Feretti et al. 1997).
The X-ray peak is shifted to the West with respect to this position.
The large field of view of the radio image shows the location
of the cluster radio galaxies with respect to the gas density distribution.  
For clarity the first radio contour is at 25$\sigma$. 
Therefore we see only the radio galaxies 
of the cluster while the low brightness diffuse emissions 
(radio halo and relic) are not visible here.

The high resolution (FWHM $\simeq$ 2\arcsec$\times$2\arcsec) images obtained at 6\,cm for the four
radio galaxies analyzed in this work are inset in Fig.~\ref{6cm}.
The sources are located at different distances from the cluster center 
which allows,
through the study of their RM (presented in Sect. 4), 
the cluster magnetic fields to be estimated along different lines-of-sight.

\begin{figure*}
\centering
\includegraphics[width=17cm]{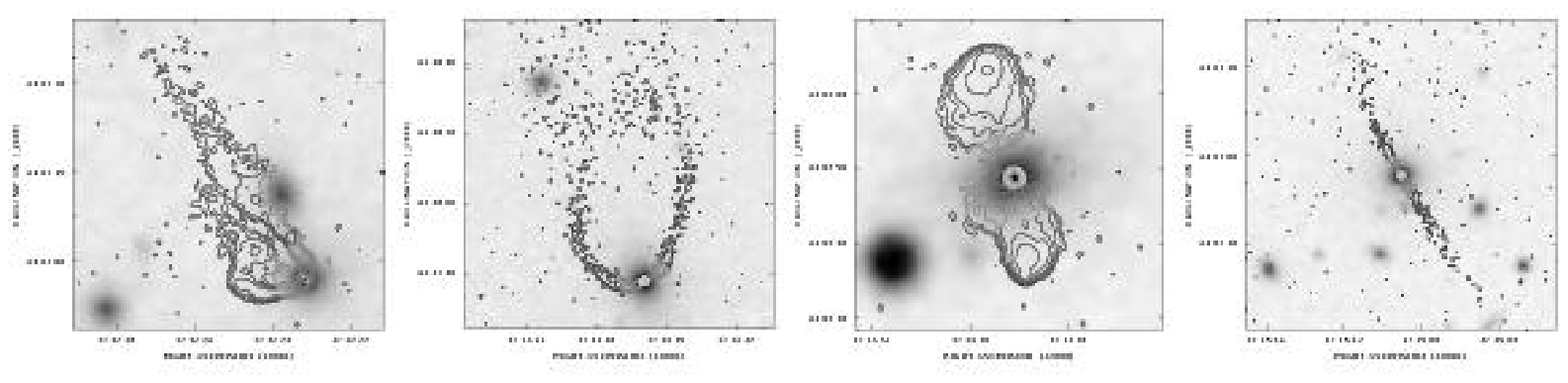}
\caption[]{
Radio contours at 3.6 cm of four cluster radio galaxies 
overlaid on the DSS2 red plate (J1712.4+6401, J1713.3+6347, J1713.5+6402 and J1715.1+6402, from left to right).
The radio contours are: 0.04, 0.08, 0.16, 0.32, 0.64, 1.28, 2.56, and 5.12 mJy/beam.
See Table \ref{mappeIhigh} for more information (e.g. resolution, sensitivity, peak brightness)
of these observations.}
\label{3cm}
\end{figure*}

Fig.~\ref{3cm} shows the high resolution (FWHM $\simeq$ 1\arcsec$\times$1\arcsec) 
observations in the 3.6 cm band
for the four sources overlaid on the DSS2 
red plate.\footnote{htpp://archive.eso.org/dss/dss}
Every source has a clear optical
counterpart and all are classified as cluster radio galaxies 
on the basis of the optical spectroscopy analysis 
(Miller \& Owen 2003; Yuan et al. 2003).

The relevant parameters of the radio images at high resolution
are listed in Table \ref{mappeIhigh}.

\begin{figure}
\centering
\includegraphics[width=9cm]{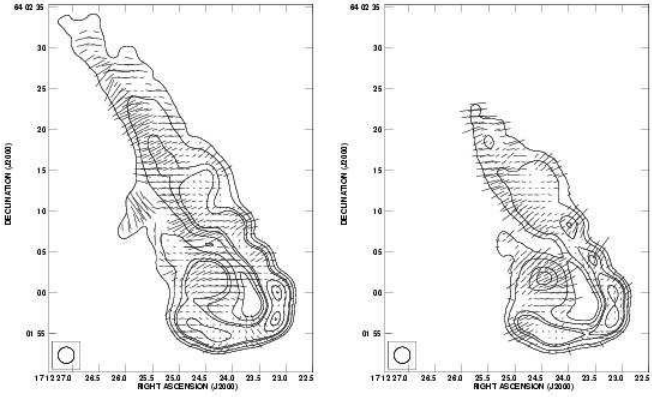}
\caption[]{Source J1712.4+6401:
Left: Total intensity contours and polarization vectors at 6 cm (4535 MHz). 
Right: Total intensity contours and polarization vectors at 3.6 cm (8465 MHz). 
The angular resolution is $2.0'' \times 2.0''$.
Contour levels are drawn at:$-$0.2, 0.2, 0.4, 0.8, 1, 2, 3, and 3.5 mJy/beam.
The lines give the orientation of the
electric vector position angle (E-field) and are proportional 
in length to the fractional polarization ($1'' \simeq16.7$\%).}
\label{a2255_b}
\end{figure}

\begin{figure}
\centering
\includegraphics[width=9cm]{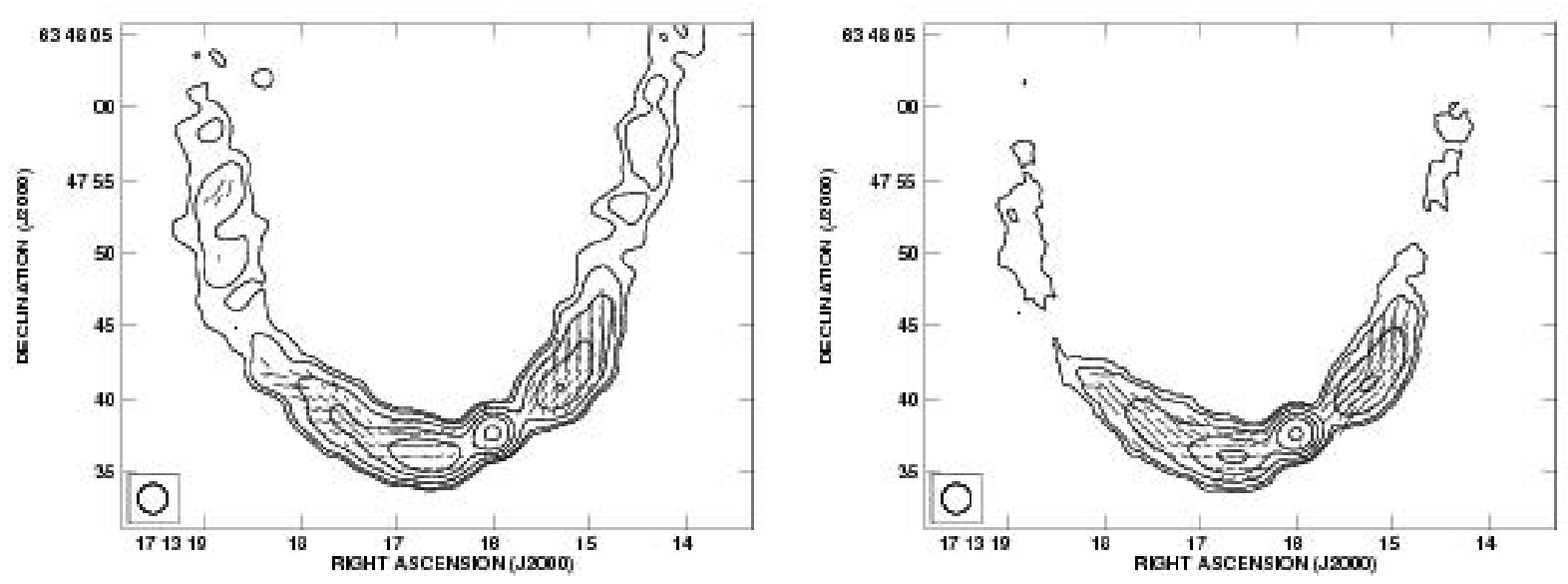}
\caption[]{Source J1713.3+6347:
Left: Total intensity contours and polarization vectors at 6 cm (4535 MHz). 
Right: Total intensity contours and polarization vectors at 3.6 cm (8465 MHz). 
The angular resolution is $2.0'' \times 2.0''$.
Contour levels are drawn at: $-$0.1, 0.1, 0.15, 0.3, 0.6, 1, and 2 mJy/beam.
The lines give the orientation of the
electric vector position angle (E-field) and are proportional 
in length to the fractional polarization ($1'' \simeq16.7$\%).}
\label{a2255_d}
\end{figure}

\begin{figure}
\centering
\includegraphics[width=9cm]{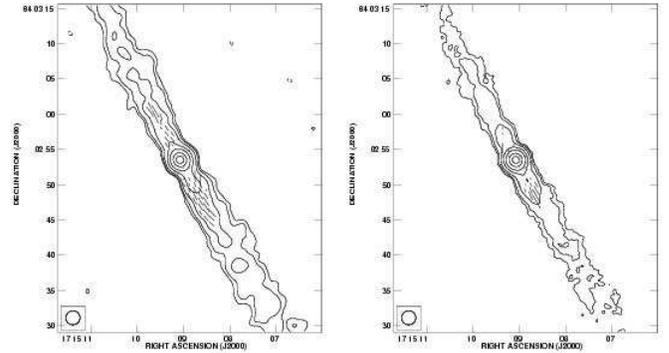}
\caption[]{Source J1715.1+6402:
Left: Total intensity contours and polarization vectors at 6 cm (4535 MHz). 
Right: Total intensity contours and polarization vectors at 3.6 cm (8465 MHz). 
The angular resolution is $2.0'' \times 2.0''$.
Contour levels are drawn at: $-$0.06, 0.06, 0.1, 0.2, 0.5, 1, 2, and 3 mJy/beam.
The lines give the orientation of the
electric vector position angle (E-field) and are proportional 
in length to the fractional polarization ($1'' \simeq16.7$\%).}
\label{a2255_f}
\end{figure}

In the following we give a brief description of the individual sources.
In the radio images presented in Figs.~\ref{a2255_b} $-$ \ref{a2255_f}
we restored the maps at different frequencies with the same 
beam (2\arcsec$\times$2\arcsec). 
The relevant parameters of these total and polarization intensity images are listed 
in Table \ref{mappeIlow}.
The intensity and polarized flux densities were 
obtained, after the primary beam correction, 
by integrating in the same area the $I$ and $P$ surface brightness, respectively, 
down to the noise level.
In Figs.~\ref{a2255_b} $-$ \ref{a2255_f} contours represent total intensity 
while vectors represent the orientation of the projected 
E-field and are proportional in length to the fractional polarization.
In the fractional polarization images, pixels with error greater than $10\%$ were blanked.
The fractional polarization values given in the following subsections are calculated in the
regions where the error is less than $10\%$. 

\begin{table}
\caption{Total intensity images at high resolution (see Figs.~\ref{6cm} $-$ \ref{3cm}).}
\label{mappeIhigh}
\centering
\begin{tabular} {c c c c c} 
\hline\hline
Source       & $\lambda$    &Beam      & $\sigma$(I)  & Peak  \\
             & (cm)         &($''$)    & (mJy/beam)   & (mJy/beam)       \\ 
\hline
J1712.4+6401 &  3.6  & 1.07$\times$0.97   & 0.014 & 1.9      \\
$''$         &  6    & 1.94$\times$1.75   & 0.019 & 3.2      \\
J1713.3+6347 &  3.6  & 1.09$\times$1.06   & 0.013 & 2.2      \\ 
$''$         &  6    & 1.99$\times$1.82   & 0.018 & 2.4      \\
J1713.5+6402 &  3.6  & 1.19$\times$1.02   & 0.014 & 11.6     \\ 
$''$         &  6    & 2.11$\times$1.82   & 0.021 & 12.8     \\ 
J1715.1+6402 &  3.6  & 1.05$\times$1.03   & 0.013 & 3.2      \\ 
$''$         &  6    & 1.82$\times$1.78   & 0.016 & 3.5      \\ \hline
\multicolumn{5}{l}{\scriptsize Col. 1: Source; Col. 2: Observation wavelength; Col. 3: Beam;}\\
\multicolumn{5}{l}{\scriptsize Col. 4: RMS noise of the I image; Col. 5: Peak brightness. }\\
\end{tabular}
\end{table}

\begin{table*}
\caption{Total and polarization intensity radio images restored with a beam of 2\arcsec (see Figs.~\ref{a2255_b} $-$ \ref{a2255_f}).}
\label{mappeIlow}
\centering
\begin{tabular} {c c c c c c c c c} 
\hline\hline
Source       & $\lambda$    &Beam      & $\sigma$(I)$^{*}$  & $\sigma$(Q)$^{*}$ & $\sigma$(U)$^{*}$ & Peak brightness  &Flux density & Pol. flux \\
             & (cm)         &($''$)    & (mJy/beam)   & (mJy/beam)  & (mJy/beam)  & (mJy/beam)         & (mJy)       & (mJy)     \\ 
\hline
J1712.4+6401 &  3.6  & 2.0$\times$2.0   & 0.015 & 0.018 & 0.017  & 2.8      & 49.5      & 9.5  \\
$''$         &  6    & $''$             & 0.019 & 0.020 & 0.020  & 3.5      & 86.0      & 13.0  \\
J1713.3+6347 &  3.6  & $''$             & 0.014 & 0.016 & 0.016  & 2.3      & 37.0      & 8.0       \\ 
$''$         &  6    & $''$             & 0.018 & 0.020 & 0.020  & 2.4      & 64.5      & 10.5  \\
J1713.5+6402 &  3.6  & $''$             & 0.014 & 0.017 & 0.018  & 11.6     & 87.0      & 10.0   \\ 
$''$         &  6    & $''$             & 0.021 & 0.024 & 0.022  & 12.7     & 123.5     & 11.5     \\ 
J1715.1+6402 &  3.6  & $''$             & 0.014 & 0.017 & 0.015  & 3.5      & 26.0      &  2.0       \\ 
$''$         &  6    & $''$             & 0.016 & 0.020 & 0.021  & 3.5      & 43.5      &  2.0       \\ \hline
\multicolumn{9}{l}{\scriptsize Col. 1: Source; Col. 2: Observation wavelength; Col. 3: Beam; 
                    Col. 4, 5, 6: RMS noise of the I, Q, U images;  }\\
\multicolumn{9}{l}{\scriptsize $^{*}$ Note that while the I images have been obtained by averaging
                    the two IFs in the same band, the Q and U images have been obtained for each}\\ 
\multicolumn{9}{l}{\scriptsize IF separately. Here we give the values of $\sigma$(Q) and $\sigma$(U) 
for the frequencies 8465 MHz and 4535 MHz at 3.6 cm and 6 cm respectively.}\\
\multicolumn{9}{l}{\scriptsize Col. 7: Peak brightness; Col. 8: Flux density; Col. 9: 
Polarized flux density (for the frequencies 8465 MHz and 4535 MHz at 3.6 cm and 6 cm respectively.)}\\
\end{tabular}
\end{table*}

\subsection{J1712.4+6401}
The source has a narrow-angle tail structure
and it is located in projection quite near to the cluster center (see Fig.~\ref{6cm}).
The images show an unresolved core,
in the position
RA(J2000)=17$^h$12$^m$23$^s$, DEC(J2000)=64\degrees01\arcmin57\arcsec,
South-West of the cluster center, and 
a long tail elongated North-East in the direction of the cluster 
X-ray centroid.
The maximum projected angular size of the source at 6 cm is 
about 70\arcsec($\simeq$105 kpc) but it appears much more extended
in our previous observation at 20 cm (2.5\arcmin).
At 3.6 cm only the brightest emission of the tail is detected.

Fig.~\ref{a2255_b} shows the polarization images of the source   
with an angular resolution of $2''\times2''$.
The source is strongly polarized both at 6 cm and 3.6 cm with 
a mean fractional polarization of about 14\% and 12.5\%
respectively.
In the core the fraction polarization is 2\% at 6 cm and 4\% at 3.6 cm 
and increases along the tail.

\subsection{J1713.3+6347}
This narrow-angle tail radio galaxy is located
in the southern part of the cluster far from the center (see Fig.~\ref{6cm}).
From the compact component, in the position 
RA(J2000)=17$^h$13$^m$16$^s$, DEC(J2000)=63\degrees47\arcmin37\arcsec~
two jets emanate and then bend to the north.
The maximum projected angular size at 6 cm is about 90\arcsec~
($\simeq$135 kpc) but the source appears much more extended at 20 cm ($\simeq$3\arcmin).
At 3.6 cm only the brightest emission of the jets have been detected.

Fig.~\ref{a2255_d} shows the polarization images of the jets
with an angular resolution of $2''\times2''$.
The sensitivity of the observations reveal significant polarization
only in the inner parts of the jets up to a distance of about 15\arcsec($\simeq$20 kpc)
from the core. 
The mean fractional polarization for the source 
is about 11\% at 6 cm and 14\% at 3.6 cm.
In the core the fraction polarization is 2\% at 6 cm and 0.4\% at 3.6 cm. 
The jet on the east shows a lower fractional polarization  
(10\% and 12\% at 6 cm and 3.6 cm respectively)
than the jet on the west (16\% and 23\% at 6 cm and 3.6 cm respectively).

\subsection{J1713.5+6402}
The source has a total extension of about 35\arcsec~($\simeq$50 kpc)
and a double structure with an unresolved core in
the position 
RA(J2000)=17$^h$13$^m$29$^s$, DEC(J2000)=64\degrees02\arcmin49\arcsec (see Fig.~\ref{6cm}). 

Fig.~\ref{a2255_e} shows the polarization images of the source
at 6 cm and 3.6 cm  with an angular resolution of $2''\times2''$.
The mean fractional polarization is $\simeq$11\% at 6 cm 
and $\simeq$13\% at 3.6 cm.
The fraction polarization of the core is 2\% at 6 cm and 3\% at 3.6 cm
while the lobes are polarized at similar levels.

\subsection{J1715.1+6402}
This wide-angle tail source, located in the periphery of the cluster (see Fig.~\ref{6cm}), 
is very extended
with a size of about 3.5\arcmin at 20 cm.
The maximum projected angular size at 6 cm is about 200\arcsec~ 
($\simeq$300 kpc) but the low surface
brightness lobes are only marginally 
detected at 6 cm and completely missed at 3.6 cm 
where only the first 30\arcsec~(45 kpc) of the jets are visible.

An unresolved component is located in the position:
RA(2000)=17$^h$15$^m$09$^s$,DEC(2000)=64\degrees02\arcmin54\arcsec.
Twin low brightness oppositely directed jets emanate from the core to the north-east
and south-west. After 40\arcsec~(60 kpc) the north-east jet
bends to the west toward the X-ray centroid while the other jet 
remains straight.

\begin{figure}
\centering
\includegraphics[width=9cm]{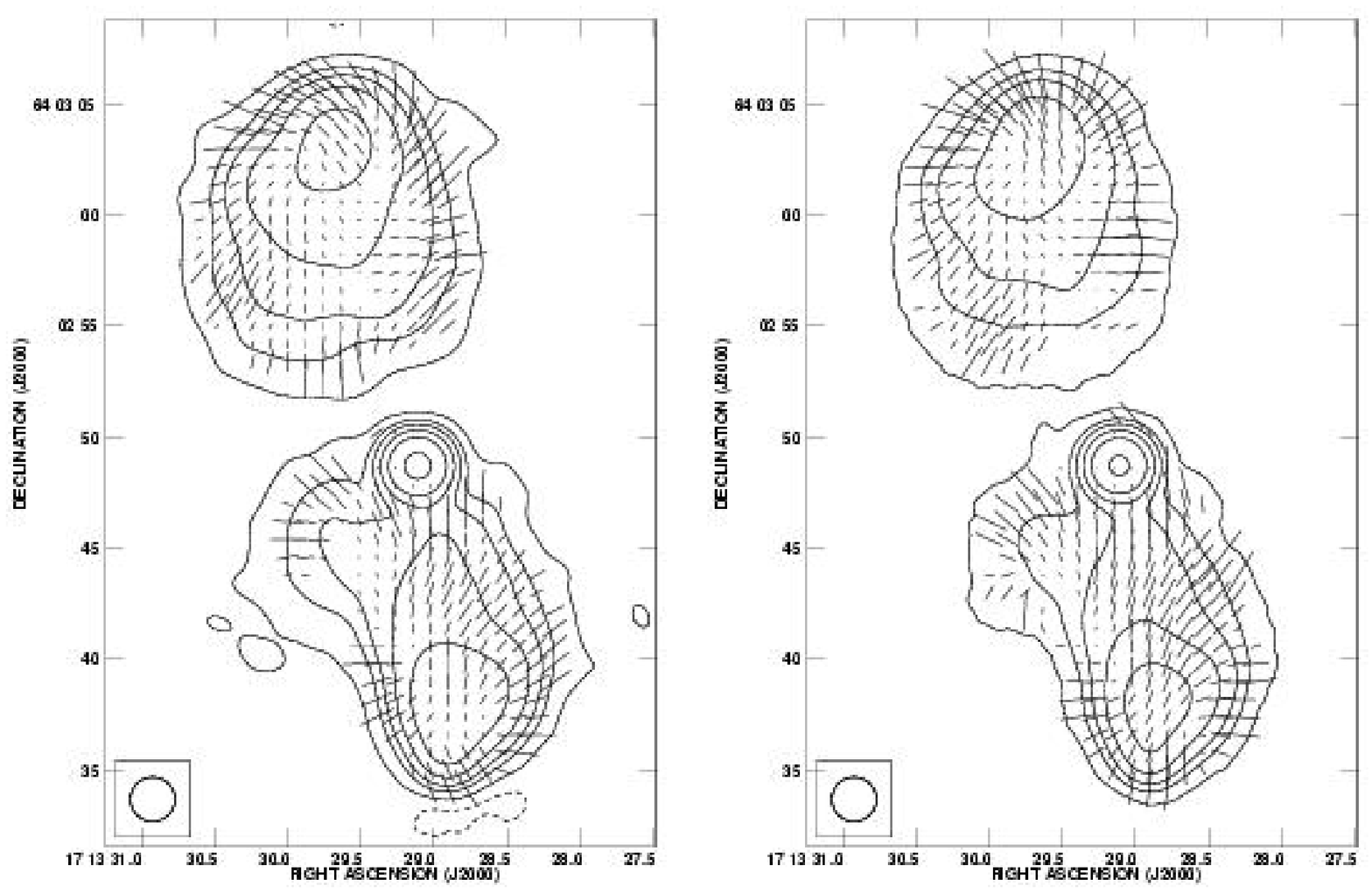}
\caption[]{Source J1713.5+6402:
Left: Total intensity contours and polarization vectors at 6 cm (4535 MHz). 
Right: Total intensity contours and polarization vectors at 3.6 cm (8465 MHz). 
The angular resolution is $2.0'' \times 2.0''$.
Contour levels are drawn at: $-$0.1, 0.1, 0.5, 1, 2, 4, and 10 mJy/beam.
The lines give the orientation of the
electric vector position angle (E-field) and are proportional 
in length to the fractional polarization ($1'' \simeq16.7$\%).}
\label{a2255_e}
\end{figure}

Fig.~\ref{a2255_e} shows the polarization images of the
central part of the source   
with an angular resolution of $2''\times2''$.
We detect polarization only in the core and in the inner arcseconds of the jets.
In the core the fraction polarization is 0.6\% at 6 cm and 2\% at 3.6 cm
while in both jets it is about 14\% at 6 cm.
The polarization in the northern jet at 3.6 cm is below the sensitivity of
the present observation while in the southern jets it is 12\%.

\section{Rotation measure images}
Polarized radiation from cluster and background radio galaxies 
may be rotated by the Faraday effect if magnetic fields are present
in the intra-cluster medium.
In this case, the observed polarization angle ($\Psi_{Obs}$)
at a frequency $\nu$
is connected to the intrinsic polarization angle ($\Psi_{int}$)
through:
\begin{equation}
 \Psi_{Obs}(\nu) = \Psi_{int} + (c/\nu)^2 \times RM 
\label{rm}
\end{equation}
where the Rotation Measure (RM) is related to the electron 
density ($n_e$), the magnetic field along the line-of-sight ($B_{\parallel}$),
and the path-length (L)
through the intracluster medium according to:
\begin{equation}
RM_{\rm~[rad/m^2]}=812\int_{0}^{L_{[kpc]}}n_{e~[cm^{-3}]}B_{\parallel~[\mu G]}dl
\label{equaz}
\end{equation}

We derived the rotation measure images, at 2\arcsec~ resolution,
of the four sources using the polarization angle maps $\Psi_{Obs}$ 
at the frequencies 4535, 4885, 8085 and 8465 MHz.
The Faraday RM images were obtained by performing a 
fit of the polarization angle images at each pixel
as a function of $\nu^{-2}$ (see Eq.\ref{rm}) 
using the algorithm
PACERMAN (Polarization Angle CorrEcting Rotation Measure ANalysis)
by Dolag et al. (2005). 
Instead of solving the n$\pi$-ambiguity
for each pixel independently, 
the algorithm solves the n$\pi$-ambiguity for a high signal-to-noise 
region and uses this information to assist computations in adjacent 
low signal-to-noise areas.

\begin{figure*}
\centering
\includegraphics[width=15cm]{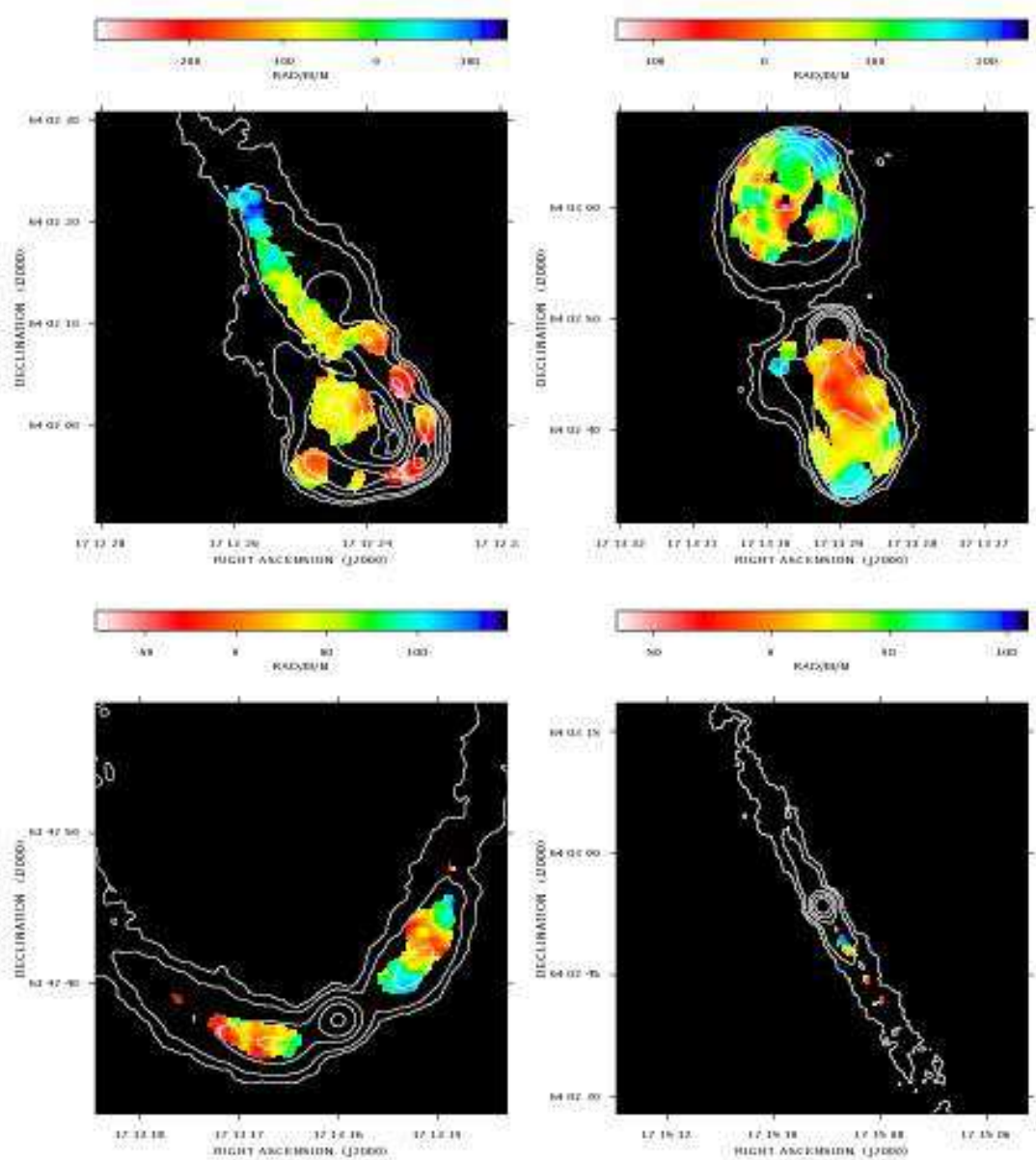}
\caption[]{Images of the rotation measure  
computed using the polarization angle maps at
the frequencies 4535, 4885, 8085 and 8465 MHz with a resolution of 2$''$.
Contours refer to the total intensity images at 3.6 cm.
Contour levels are: 0.06, 0.15, 0.5, 1, and 2 mJy/beam.}
\label{rmfig}
\end{figure*}

\begin{figure}[t]
\centering
\includegraphics[width=8cm]{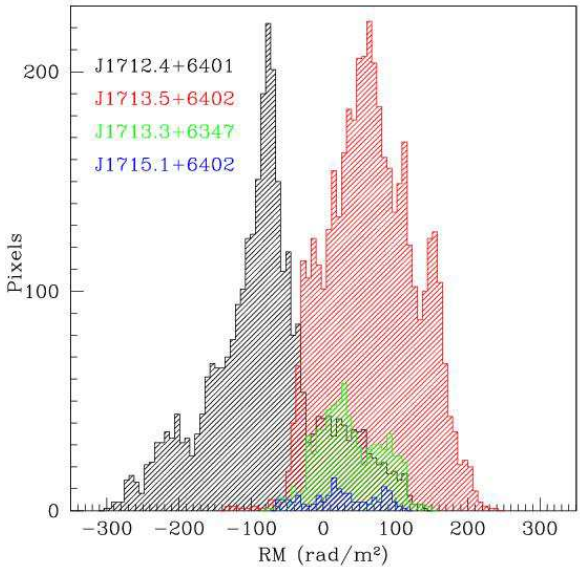}
\caption[]{Histograms of the rotation measure images for all significant pixels.}
\label{rmhist}
\end{figure}

Fig.~\ref{rmfig} shows the total intensity contours at 3.6 cm
 overlaid on the RM images of the four cluster galaxies.
The RM values range from about $-$300 rad/m$^2$ up to 250 rad/m$^2$
and reveal patchy structures with RM fluctuations 
down to scales of a few kpc.
Fig.~\ref{rmhist} shows the 
histograms of the RM distribution for the four sources.
Due to the very low number of reliable RM pixels in J1715.1+6402
this source will not be considered in the following statistical analysis.
Table \ref{rmtab} reports, 
for the other three sources sorted by distance from the X-ray
centroid, the mean value \rmm\, the
root mean square \srm\ and the maximum absolute value $\arrowvert {\rm RM_{max}} \arrowvert$ of the RM distribution.
These data are not corrected for the Galactic contribution 
which is likely negligible. A2255
in galactic coordinates is located at lon=93.97\degrees and 
lat=$+34.95$\degrees~ and based on the average RM for extragalactic
sources published by Simard-Normandin et al. (1981),
the RM Galactic contribution in the region occupied by A2255 is expected to be
about $-6$~rad/m$^2$.

Our data suffer from two kinds of uncertainty.
The first uncertainty is related to the fit 
and takes into account the presence of errors in the measurements.
In the RM images all the pixels with an error in the fit greater 
than 40 rad/m$^2$
were blanked. This ensures that at all 
frequencies we always consider pixels with an error
in polarization angle lower than 10\degrees~ (see Eq. \ref{rm}). 
The mean fit error calculated in the images
is about 20 rad/m$^2$ for J1713.5+6402 and about 30 rad/m$^2$ for J1712.4+6401 
and J1713.3+6347.
The second source of error is the statistical error.
Ideally one would determine the parameters of the RM distribution
by investigating an area as large as possible.
The morphology of sources and the presence of blanked pixels in the RM
images introduce a further uncertainty 
in this determination.
This uncertainty is particularly strong in sources
where the reliable RM pixels are present in a small 
number of observing beams $N$ (see Table \ref{rmtab}). The statistical error has been 
calculated to be about 10 and 5 rad/m$^2$ for the 
\rmm~ and \srm~respectively in J1713.5+6402 and 
about 15 and 10 rad/m$^2$ for the \rmm~ and \srm~ 
respectively in the other two sources.

As found in other studies (Feretti et al. 1999,
Taylor et al. 2001, Govoni et al. 2001a) aimed to study the cluster magnetic
field by sampling the RM of extended radio galaxies located in the same 
cluster of galaxies, the most striking result is the trend of
the RM values with distance from the cluster center. 
The innermost source, J1712.4+6401, has the highest \srm~ and \absrmm. 
The source J1713.5+6402, located at about 
1 core radius from the cluster center, 
shows considerably lower RM values. Finally, the peripheral source J1713.3+6347
shows a \srm~ consistent with its uncertainty but still has a significant \rmm.

The RM results are in agreement with the interpretation
that the external Faraday screen is the same 
for all the sources, i.e. the decreasing radial profile
of the RM data may be due to the intracluster medium of A2255,
whose differential contribution depends on how much magneto-ionized 
medium is crossed by the polarized emission.
Thus the data are consistent with the existence of a magnetic field 
associated with the intracluster medium.

The RM structures on small scales can be explained by the fact
that the cluster magnetic field fluctuates on scales
smaller than the size of the sources. On the other hand,
the RM distribution with a non-zero mean indicate
that the magnetic field fluctuates also on scales larger than the radio sources.
Therefore to study in detail the cluster magnetic field properties
it seems necessary to consider cluster magnetic fields 
models where both small and large scale coexist.

\begin{table*}
\caption{Rotation measure.}
\label{rmtab}
\centering
\begin{tabular} {c c c c c c} 
\hline\hline
Source   & Centroid Dist. & $<{\rm RM}>$         &   $\sigma_{RM}$  & $\arrowvert {\rm RM_{max}}\arrowvert$ & N \\
         &  (kpc)         & (rad/m$^{2}$)  &   (rad/m$^{2}$)  & (rad/m$^{2}$)       & (Beams)\\
\hline 
J1712.4+6401  & 315  & $-$81        &   79 & 300& 29  \\
J1713.5+6402  & 450  &    67        &   59 & 236& 41  \\ 
J1713.3+6347  & 1570 &    36        &   42 & 149& 6   \\ \hline 
\multicolumn{5}{l}{\scriptsize Col. 1: Source; Col. 2:  Projected distance from the X-ray centroid; Col. 3: Mean of the RM distribution;}\\
\multicolumn{5}{l}{\scriptsize Col. 4: RMS of the RM distribution. Col. 5: Maximum absolute value of the RM distribution; }\\
\multicolumn{5}{l}{\scriptsize Col. 6: Number of independent beams in the RM images.}\\
\end{tabular}
\label{rmtab}
\end{table*}

\section{The FARADAY tool}

An analysis of the rotation measure 
of radio sources sampling different lines-of-sight across the cluster,
together with an X-ray observation of the intracluster gas,
can be used to derive
information on the strength and 
structure of the cluster magnetic field.
Complementary information on the cluster magnetic field
can be derived by analyzing the radio halo emission.
In particular, for a given distribution of the
relativistic electrons in a cluster, different magnetic field models
will generate very different total intensity and polarization
brightness distributions for the radio halo emission.
For those clusters containing a radio halo, a realistic magnetic field 
modeling should be able to explain both the small scale fluctuations seen 
in the RM images of cluster radio galaxies, and the morphology 
and polarization of the large scale diffuse emission.  

The FARADAY tool, described in Murgia et al. (2004), 
permits an investigation of 
cluster magnetic fields by comparing the observations
with simulated RM and radio halo images, obtained by considering 
3-dimensional multi-scale cluster 
magnetic field models.
We applied this method to A2255 with the aim of finding the optimal
magnetic field strength and structure capable of describing 
both the high resolution RM images presented in this work 
and the large scale polarized features visible in the radio halo of A2255
(Govoni et al. 2005).

There are a number of quantities that are critical in our modeling, these are:
 the gas density distribution of the thermal electrons, 
the characterization of the magnetic field fluctuations
and radial scaling, the energy density distribution of 
the synchrotron electrons. 
 
For the distribution of the thermal electron gas density in A2255
we assumed a standard $\beta$-model profile:
\begin{equation}
n_{\rm e}(r)=n_{\rm 0}(1+r^2/r^2_{\rm c})^{-3\beta/2}
\label{king}
\end{equation}
\noindent
where $r$, $n_{\rm 0}$ and $r_{\rm c}$ are the 
distance from the cluster X-ray centroid, the central electron density, 
and the cluster core radius, respectively.
The gas density distribution
has been taken as derived from the ROSAT X-ray 
observations of Feretti et al. (1997), rescaled to our chosen cosmology
($r_{\rm c}$=432 kpc, $n_{\rm0}=$2.05$\times$10$^{-3}$ 
cm$^{-3}$, $\beta$=0.74).

We considered different power law cluster magnetic field power spectrum\footnote{ Note that throughout this 
paper the power spectra are expressed as vectorial forms in $k$-space. 
The one-dimensional forms can be obtained by multiplying by $4\pi k^{2}$ and $2\pi k$
 respectively the three and two-dimensional power spectra. According to this notation the Kolmogorov 
spectral index is $n=11/3$.} 
models:
\begin{equation}
|B_k|^2\propto k^{-n}
\end{equation}
\noindent
in a 3-dimensional cubical box.
We used a grid of 1024$^{3}$ pixels in size which
has the necessary resolution to sample the
magnetic field power spectrum on a spatial scale\footnote{Here we refer to 
the length $\Lambda$ as the magnetic field reversal scale. In this way, $\Lambda$ 
corresponds to a half-wavelength, i.e. $\Lambda= 0.5\cdot(2\pi/k)$.} 
ranging from $\Lambda_{\rm
min}=4$\,kpc up to $\Lambda_{\rm max}=512$\,kpc. 
Observations reveal RM fluctuations on small scales 
supporting the choice for $\Lambda_{\rm min}$.
The exact value of $\Lambda_{\rm max}$ is more uncertain, however
the presence of ordered polarized filaments in the radio halos of A2255  
indicates that, in the case of this cluster, the magnetic field fluctuates up 
to such large scales. 

We adopted a magnetic field distribution decreasing from the cluster center according to:
\begin{equation}
\langle\mathbf B\rangle(r)=\langle\mathbf B\rangle_{\rm 0}\cdot(1+r^{2}/r_{\rm c}^{2})^{-3\mu/2}
\label{br}
\end{equation}
\noindent
where $\langle\mathbf B\rangle_{\rm 0}$ is the mean magnetic field at the cluster center.
In order to estimate the index $\mu$, we first compared the \srm~and the intracluster X-ray surface 
brightness, $S_{\rm X}$, at the location
 of each radio galaxy (Dolag et al. 2001). In the case of a beta model, 
the index $\mu$ is related to the slope, $\alpha$, of the $\sigma_{\rm RM}\propto S_{\rm X}^{\alpha}$ 
correlation through $\mu=(2\alpha-1)\cdot(\beta-1/6)$,
see also Murgia et al. (2004). In the case of the three radio galaxies in A2255 the fit of the $\sigma_{\rm RM}- S_{\rm X}$ correlation yields $\alpha=0.6\pm 0.5$. The value of $\mu$ obtained with this method is indeed rather uncertain and lies in the range from $-$0.4 to 0.7.
We thus estimated $\mu$ by comparing $S_{\rm X}$ and radio halo surface brightness at 
 increasing radial distance from the cluster center (Govoni et al. 2001b). In A2255 it is observed a
 linear correlation between these two quantities,
indicating that thermal and non-thermal (relativistic particles and 
magnetic fields) energy densities could have the same
radial scaling. Therefore in the simulations we adopted $\mu=\beta/2$=0.37 
which corresponds to a magnetic field whose energy density 
decreases in the same way as the gas energy density.

Provided the distribution of the thermal electrons and the magnetic field model
described above we simulated the RM by integrating Eq.\,\ref{equaz} along the line-of-sight.

To simulate the radio halo emission 
at each point of the computation grid we calculated the 
synchrotron emissivity by convolving the emission spectrum 
of a single relativistic electron with the particle energy distribution
of an isotropic population of
relativistic electrons whose distribution follows:
\begin{equation} 
N(\epsilon,\theta)=N_{\rm 0}\epsilon^{-\delta}(\sin\theta)/2 
\label{N}
\end{equation}
where $\epsilon$ and $\theta$ are the electron's energy and the pitch 
angle between the electron's velocity and the direction of the 
magnetic field, respectively (see also Murgia et al. 2004).

The energy density of the relativistic electrons is:
\begin{equation} 
  u_{\rm el}=\int_{\epsilon_{\rm min}}^{\epsilon_{\rm max}}N(\epsilon)\epsilon d\epsilon 
\label{uel_integral}
\end{equation}
where $\epsilon_{\rm min}$ and  $\epsilon_{\rm max}$ are the low and high energy cut offs
 of the energy spectrum, respectively.

In our model the  magnetic field energy density $ u_{\rm B}=B^2/8\pi$ and $u_{\rm el}$ 
are in equipartition at every 
point in the cluster, therefore both energy densities have the same radial decrease.
We adopted an electron energy spectral index
$\delta=3$ and a Lorentz factor for the
high-energy cutoff of the electron distribution $\gamma_{\rm
max}=1.5\times 10^{4}$. The resulting radio halo emission spectrum is
roughly consistent with the radio spectral index map (not shown, Govoni et al. in preparation) 
obtained by combining our radio halo image at 1.4 GHz (Govoni et al. 2005) with
the image at 327 MHz (Feretti et al. 1997).
With $\delta=3$ we have $u_{\rm el}\simeq N_{\rm 0}/\gamma_{\rm min}$.
The low-energy cutoff $\gamma_{\rm min}$ and $N_{\rm 0}$,  
are adjusted to guarantee the observed halo brightness and $u_{\rm el}= u_{\rm B}$.
In practice we fix $\gamma_{\rm min}$=10 and let $N_{\rm 0}$ vary.

Below we present the simulated RM and radio halo images 
corresponding to two different variants of the above model. As a first 
approximation, we keep the power spectrum spectral index $n$ constant 
throughout the entire cluster (Sect.\ref{constantn}). 
However, we find that this model does not reproduce the radio halo polarization level 
observed in A2255.
To describe both the RM profiles and the radio halo polarization
the data require a steepening of the power spectrum spectral index
from the cluster center to the periphery and the presence of filamentary structures
on large scales (i.e. a non-Gaussian magnetic field).
In Sect.\ref{variablen} we will show that a model with these characteristics
can better account for both the observed RM statistic and the radio
halo polarization.

\begin{table}[]
\caption{Gas density and magnetic field parameters adopted in the simulations.}
\label{simul}
\centering
\begin{tabular}{l l}
\hline
\hline                  
 Grid size$^{a)}$       & $1024^{3}$ pixels \\   
 Cellsize               & 1 pixel=2 kpc \\
\hline
 Core radius          & $r_{\rm c}$=432 kpc         \\
 Central density      & $n_{\rm 0}$=2.05$\times$$10^{-3}$ cm$^{-3}$    \\
 Beta                 & $\beta$=0.74 \\ \hline
 Magnetic field model & $|B_k|^2\propto k^{-n}$ \\
 Magnetic field radial profile slope& $\mu=0.37$            \\
 Magnetic field minimum scale    & $\Lambda_{\rm min}$=4 kpc \\
 Magnetic field maximum scale    &  $\Lambda_{\rm max}$=512 kpc\\
 Central Mean magnetic field   & free parameter \\
 Power spectrum spectral index & free parameter (n=2-4)\\
\hline
\multicolumn{2}{l}{\scriptsize{ a) To extend simulated RM images up to the source J1713.3+6347}}\\
\multicolumn{2}{l}{\scriptsize{ the computational grid has been replicated at boundaries realizing a}}\\
\multicolumn{2}{l}{\scriptsize{field of view of 4096$\times$4096 kpc$^2$}}\\
\end{tabular}\end{table}

The computational grid, gas density, and magnetic field model
parameters of the simulations are provided in Table \ref{simul}.
A grid size of $1024^3$ pixels with a cellsize of 2 kpc 
allows us to simulate a cluster field of view of 
2048$\times$2048 kpc$^2$. The simulated 
magnetic field is periodic at the grid boundaries. Thus, to extend our RM 
simulations up to the source J1713.3+6347, the computational grid has been 
replicated realizing a field of view of 4096$\times$4096 kpc$^2$ 
(i.e. up to about one virial radius).

\section{Constant magnetic field power spectrum slope}
\label{constantn}

\subsection{Simulated Rotation Measure images} 

\begin{figure*}
\centering
\includegraphics[width=18cm]{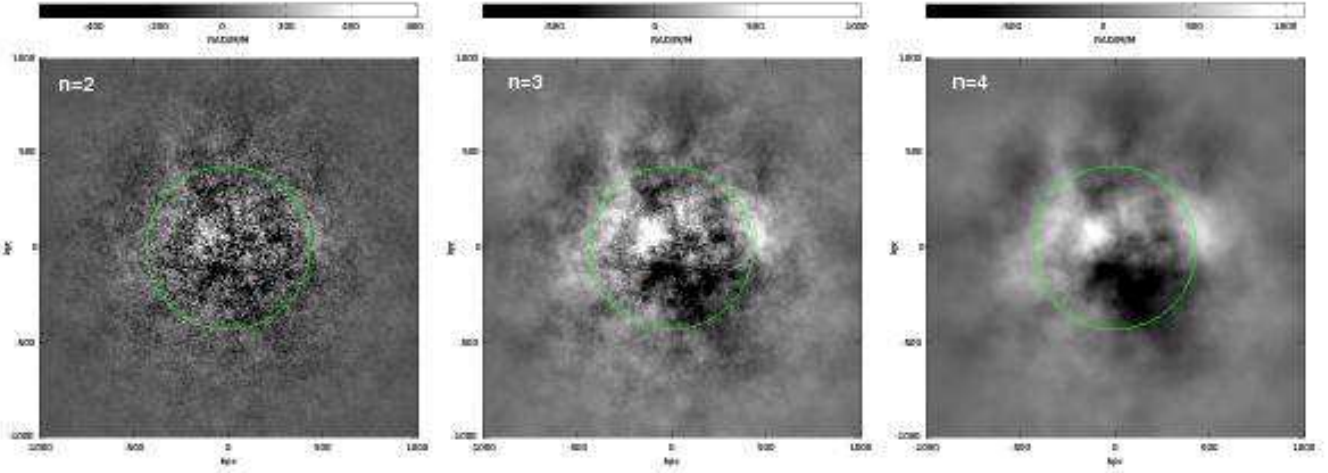}
\caption[]{
Simulated RM images for different values of the magnetic field power 
spectrum spectral index $n$. The three power spectra are normalized to have the 
same total magnetic field energy which is distributed over the range of spatial scales
from 4 kpc up to 512 kpc. 
In these simulations the average field at the cluster center is \bm$_{\rm0}$ $=2 ~\mu$G and its  
energy density decreases from the cluster center according to $B^{2}\propto n_{\rm e}(r)$.
Each RM image shows a field of view of about $2\times 2$ Mpc. 
We performed the RM integration  
from the cluster center up to 4 Mpc along the line-of-sight.
The circles represent the cluster core radius.}
\label{fig_rm}
\end{figure*}

In Fig.~\ref{fig_rm} we show the central $2\times 2$ Mpc$^2$ region of the simulated RM images 
of A2255 for three different values of the magnetic field power spectrum spectral index, $n$=2, 3 and 4.

To obtain such images we performed the integration of
Eq.~\ref{equaz} from the cluster center up to 4 Mpc 
along the line-of-sight using the magnetic field model described 
in the previous section.
The three magnetic field power spectra are normalized to generate the same
average magnetic field energy density. 
The average magnetic field strength at the cluster 
center is \bm$_{\rm 0}=2\,\mu$G and 
its energy density decreases from the cluster center following the gas energy profile. The central 
 magnetic field strength of this set of simulations provides the best fit to the observed RM radial 
profiles (see below).
It is evident from Fig.~\ref{fig_rm} that the same cluster 
magnetic field energy
density generates different magnetic field configurations with 
correspondingly different RM structures, for different values of the power
spectrum spectral index.

\begin{figure}
\begin{center}
\includegraphics[width=8cm]{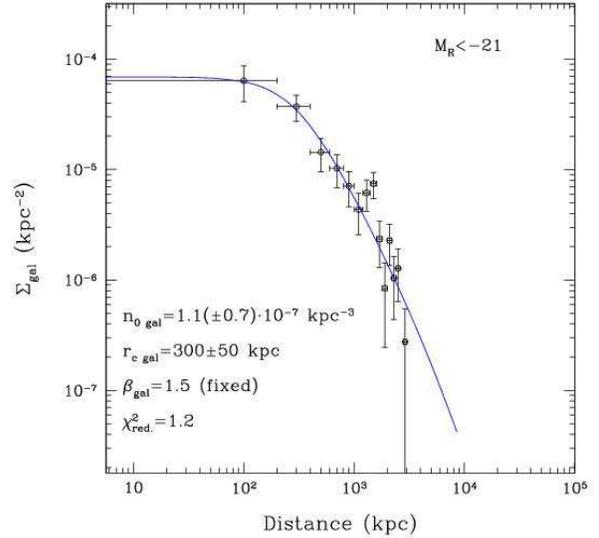}
\end{center}
\caption[]{Spatial distribution of member galaxies in A2255 taken from Yuan et al. (2003).
The solid line represents the fit of the King model $n_{\rm gal}=n_{\rm 0\,gal}(1+r^2/r_{\rm c\,gal}^2)^{-\beta_{\rm gal}}$ 
to the data. We derived a central density of galaxies $n_{\rm 0\,gal}=1.1\pm0.7\times 10^{-7}$ kpc$^{-3}$
and a core radius $r_{\rm c\,gal}=300 \pm50$ kpc. The $\beta_{\rm gal}$ 
parameter have been fixed to 3/2. 
The fit, has been extended 
only to those members characterized by a red magnitude brighter than $M_{\rm R} <-21$.
}
\label{NGAL}
\end{figure}

RM images of the whole cluster, such as those presented in 
Fig.~\ref{fig_rm}, cannot be observed in reality.
However it is possible to derive the RM from limited regions of the cluster
by observing radio sources located at different projected distances from 
the center as described in Sect. 4. 
The comparison of the observations with the simulations 
can constrain the magnetic field strength and power spectrum. 
In such a comparison the details of the three-dimensional structure of the radio sources are
neglected, and the entire observed Faraday rotation is assumed to
occur in the intracluster medium. The three radio galaxies considered
in this work are cluster members. For what concerns their location
along the line-of-sight, we followed two approaches.
As first approximation, we supposed that the three galaxies
lie on a plane perpendicular to the line-of-sight at the distance of the cluster center.
The simulations shown in Fig.~\ref{fig_rm} reproduce such a situation.
As an improvement, we extracted the Faraday depth of the simulated RM images randomly from the
3 dimensional distribution of bright member galaxies observed in A2255. We determined 
this trend by fitting with a King model the spatial galaxy distribution of the
 brightest spectroscopically confirmed member galaxies taken from Yuan et al. (2003). 
The fit, shown in  Fig.\,\ref{NGAL}, has been extended only to those members 
characterized by a red magnitude brighter than $M_{\rm R} <-21$
 since radio galaxies usually populate this optical luminosity range (e.g. Govoni et al. 2000).

To compare directly the simulations with the observations  
we added to the simulated RM images a noise of 30 rad/m$^2$  
(corresponding to the typical fit error in our data) 
and an offset of $-6$ rad/m$^2$ 
(representing the RM galactic contribution calculated in
the direction of A2255).
Therefore, in the peripheral regions of the simulated RM images, 
where the contribution of the cluster Faraday screen is negligible, 
 we have \srm$\simeq$30\,rad/m$^2$ and \rmm$\simeq-6$\,rad/m$^2$. 
These are the lower limits of our simulations.

\begin{figure*}
\centering
\includegraphics[width=12cm]{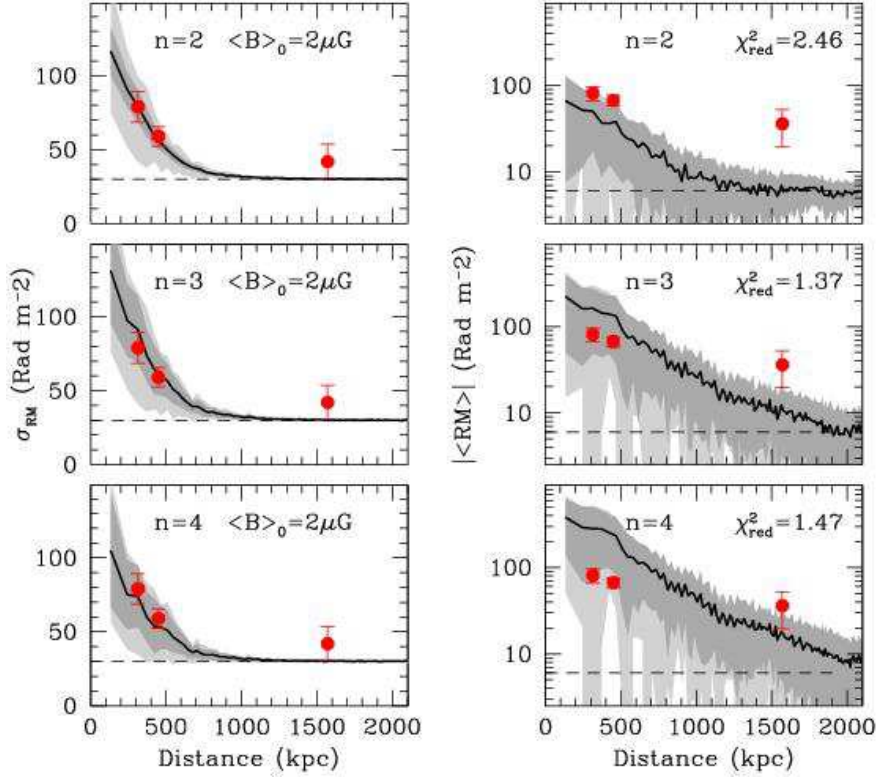}
\caption[]{
Comparison of \srm~(left column panels)
and \absrmm~(right column panels) data of the three cluster 
radio galaxies (dots) with the simulated 
profiles (dark lines). The dark gray regions
show the dispersion of the profiles calculated in the simulations shown in Fig.~\ref{fig_rm}.
The light gray regions represent the increase of the scatter of simulated RM in the case in which
the Faraday depth of the simulated RM images is extracted randomly 
from the 3 dimensional distribution of galaxies in A2255.
The horizontal dashed lines represent the lower limits of \srm~ and \rmm~
caused by the presence of fitting errors and the RM galactic contribution. In the right panels 
 the value of the reduced $\chi^{2}$ for two degree of freedom is also indicated. 
}
\label{fit_rm}
\end{figure*}

In Fig.~\ref{fit_rm} we compare the observed \srm~and \absrmm~
with the expectation of the simulations 
(for $n=2,3,4$ and \bm$_{\rm0}$ $=2 ~\mu$G, see below).
The plotted data are those presented in Table \ref{rmtab}. The error bars 
represent only the statistical errors discussed in Sect. 4 and
do not account for fitting errors since those are already  included in the simulated RM images.
Since the observed \srm~ and \rmm~ are calculated in limited regions over 
the radio galaxies, we have to sample the corresponding 
simulated values in regions of equivalent size.
The expected radial trends have been obtained 
by covering the simulated RM images with a grid of rectangular boxes 
of $50\times 50$ kpc$^{2}$ in size.
These boxes reproduce the regions covered by the RM images of the radio galaxies 
in A2255.
Inside each box we calculated the values 
of \srm~and \absrmm~exactly as it would be done if they were 
real radio sources. 
Due to the random nature of the simulated magnetic field, the values of 
\srm~and \absrmm~at a given radial distance vary from box to box. 
However, the numerical simulations allow us to quantify this 
statistical variance. 
We calculated the mean and the dispersion of \srm~and \absrmm~values from all 
the boxes located at the same projected
distance from the cluster center.
The dark lines show the mean while the dark gray regions
show the dispersion of the profiles calculated in the simulations shown in Fig.~\ref{fig_rm}.
The light gray regions represent the increase of the scatter of simulated RM in the case in which
 the location of the boxes along the line-of-sight is extracted randomly 
from the 3 dimensional distribution of galaxies in A2255.
The horizontal dashed lines represent the lower limits of \srm~ and \rmm~
caused by the presence of fitting errors and the RM galactic contribution.

Both the values of \srm~and \absrmm~ depend linearly
on the cluster magnetic field strength.
The best fit of the simulated \srm~ radial profiles with 
respect to the \srm~
values observed in A2255 was obtained by considering a central magnetic field 
of about 2\,$\mu$G. It reproduces quite well the observed \srm~ for each of the
considered values of the magnetic field spectrum spectral index ($n$=2,3,4), as shown
in the left panel of Fig.~\ref{fit_rm}.
The comparison between simulated and observed \absrmm~ 
(Fig.~\ref{fit_rm}, right panels) allows us to constrain the value of the best 
index of the power spectrum. In order to quantify the relative goodness of the fit 
 of the \absrmm~profiles we calculated the sum 
\begin{equation}
\chi^{2}=\Sigma\frac{({\arrowvert\langle{\rm RM}\rangle\arrowvert}_{\rm Sim}-{\arrowvert\langle{\rm RM}\rangle\arrowvert}_{\rm Obs})^{2}}{{\rm Scatter}_{\rm Sim}^{2}+{\rm Err}_{\langle{\rm RM}\rangle}^{2}}
\label{chi2}
\end{equation}
where ${\rm Scatter}_{\rm Sim}$ and ${\rm Err}_{\langle{\rm RM}\rangle}$ represent the scatter of the simulations
 and the statistical \absrmm~error, respectively.

A flat spectral index power spectrum, $n=2$, is ruled out since it predicts
 too low \absrmm~ for all three sources. A steep spectral index power spectrum, $n=4$, 
 fits the outer source but predicts too much \absrmm~ for the two central sources. 
The model with $n=3$ agrees better with the data.

\subsection{Simulated radio halo images}

We simulated the expected total intensity and polarization brightness
distribution at 1.4 GHz, for a magnetic field strength
$\langle B\rangle_{\rm 0}=$2\,${\rm \mu G}$ as found in the RM data.

\begin{figure*}
\centering
\includegraphics[width=18cm]{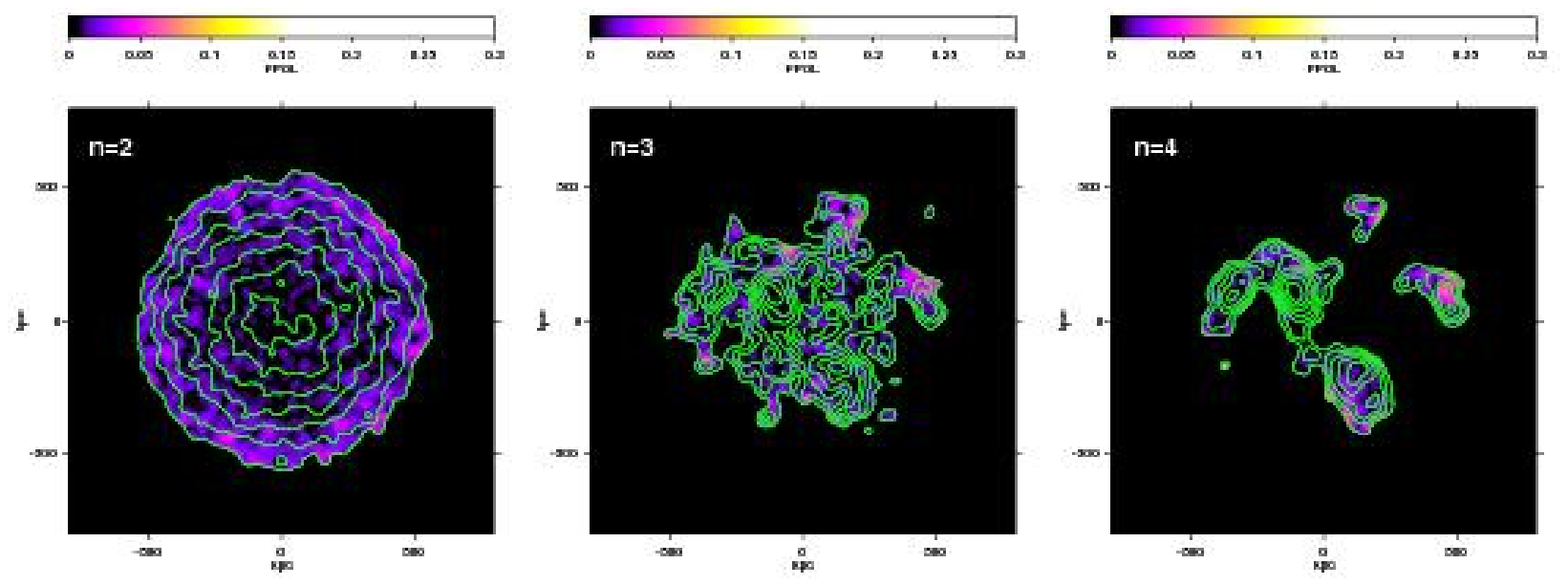}
\caption[]{
Simulated radio halo brightness
(contours) and fractional polarization (color) 
for the three different magnetic field spectral indices $n$=2, 3 and 4. 
For a direct comparison with the data (Fig.~\ref{halo_n2-4}; 
 bottom right panel)
these images have been convolved with a beam of 25\arcsec~ 
(37.5 kpc) and the lowest contour level has been chosen to match 
the dynamic range of 10 as in the observations.
}
\label{halo}
\end{figure*}

In Fig.~\ref{halo} we show the simulated radio halo brightness
 (contours) and fractional polarization (color) 
for the three different magnetic field spectral indices $n$=2, 3 and 4. 
For a direct comparison with the data (see Fig.~\ref{halo_n2-4} ; bottom right panel)
 these images have been convolved with a beam of 25\arcsec~(37.5 kpc) 
and the lowest contour level has been chosen to match 
the dynamic range of 10 as in the observations.

As in the case of the simulated RM images, the same cluster magnetic field energy
density generates different magnetic field configurations with 
correspondingly different halo morphologies, for different values of the power
spectrum spectral index, $n$. However, in this case these 
differences are even more evident given that the radio halo intensity depends
 upon the square of the magnetic field fluctuations.

Flat power spectrum indexes ($n=2$) give rise to a regular and smooth radio
 halo morphology. Increasing the spectral index, $n$, and thus the power on
 large scales, the radio halo becomes increasingly irregular. For $n=4$ 
 the radio halo emission is confined to a few bright filaments. Again, the simulated 
radio halo corresponding to $n=3$ seems to better reproduce the observed morphology.

The internal depolarization of the radio halo emission is stronger where the 
 RM is higher and where the magnetic field is tangled on small scales.
Thus, as one would expect, we found that the degree of fractional polarization
increases with increasing values of $n$ and decreases towards the cluster center (Murgia
 et al. 2004). In any case, one can observe significant polarized emission only
 from the outer layers of the halo observed along the line of sight paths that do not cross the cluster 
center.
The maximum fractional polarization attained in this set of simulations at
the outer halo edges ranges from 5\%, for $n=2$, up to 9\%, for $n=4$. These values 
are definitely lower than the values observed in the radio halo of A2255. 

Therefore,  our results seem to indicate that a single
magnetic field power spectrum slope cannot account for both the observed RM 
and radio halo images. Even $n=3$, which gives a good results 
for the RM profiles
and radio halo morphology, yields insufficient polarization levels.

\section{Variable magnetic field power spectrum slope}
\label{variablen}

\begin{figure*}
\centering
\includegraphics[width=18cm]{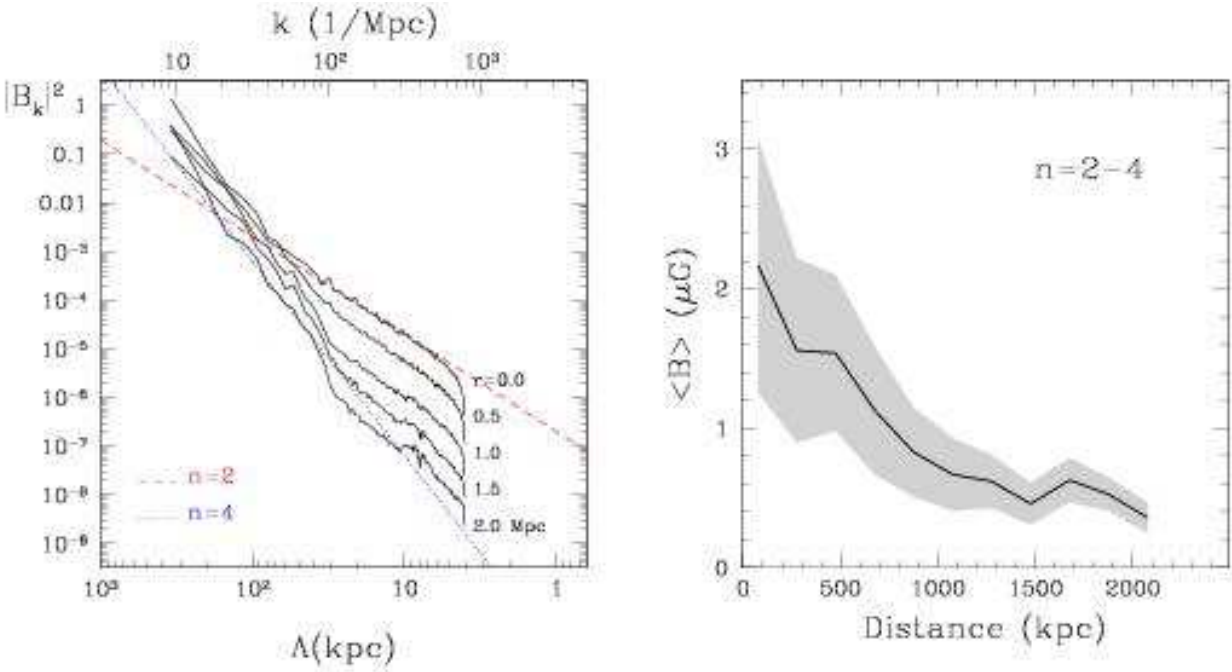}
\caption[]{
Left: magnetic field power spectra as a function of the distance from the cluster center for
the variable magnetic field power spectrum slope model. The dashed and dotted reference lines
 have a slope of 2 and 4, respectively. Right: radial profile of the magnetic field strength. The solid
 line represents the average magnetic field strength while the gray region indicates the root mean square of the
 magnetic field fluctuations.
}
\label{power}
\end{figure*}

The results of the previous section suggest that the magnetic field 
power spectrum slope may change as a function of distance from the 
cluster center. The data require that much of the magnetic field energy 
density at the cluster center should be concentrated on small scales
in order to account
 for the observed RM dispersion. On the other hand, to explain the highly
 polarized structures observed in the radio halo it is necessary that 
 most of the magnetic field energy density is on large scales at the cluster periphery.
This scenario is also consistent with the relatively large \rmm~of the most external radio galaxy
  J1713.3+6347 (see Fig.~\ref{fit_rm}).
We thus considered a magnetic field configuration whose power spectrum gradually steepens with increasing 
distance from the cluster center (see Fig.~\ref{power}).

After a series of simulations, we found that the best global magnetic field is composed by the sum of:\\
 i) a Gaussian, $n=2$, magnetic field whose intensity decays exponentially beyond the core radius 
and\\ 
ii) a non-Gaussian, $n=4$, magnetic field with $\Lambda_{\rm min}=32$\,kpc 
and $\Lambda_{\rm max}=512$ 
whose intensity decays exponentially for $r<r_{\rm c}$. The values of $\Lambda_{\rm min}$ and 
$\Lambda_{\rm max}$  adopted for the outer field are roughly consistent with the
 observed width and length of the polarized radio halo filaments, respectively.
We achieved the non-Gaussianity of the outer magnetic field fluctuations by exponentiating
a 'parent' Gaussian magnetic field with $n=4$.

Non-Gaussian fields are characterized
 by a lower  filling factor with respect to Gaussian random fields since magnetic field energy
 is more concentrated in filaments. The non-Gaussianity is indeed a necessary feature of the 
 outer field since it allow us to enhance the field strength in the polarized filaments 
boosting their prominence. Moreover, the presence of large voids in the magnetic field 
 ensures that the \rmm~ over the two central radio galaxies is not exceeded as would happen
 for a Gaussian field with a similar power spectrum slope (see bottom left panel of Fig.~\ref{fit_rm}).

\begin{figure*}
\centering
\includegraphics[width=12cm]{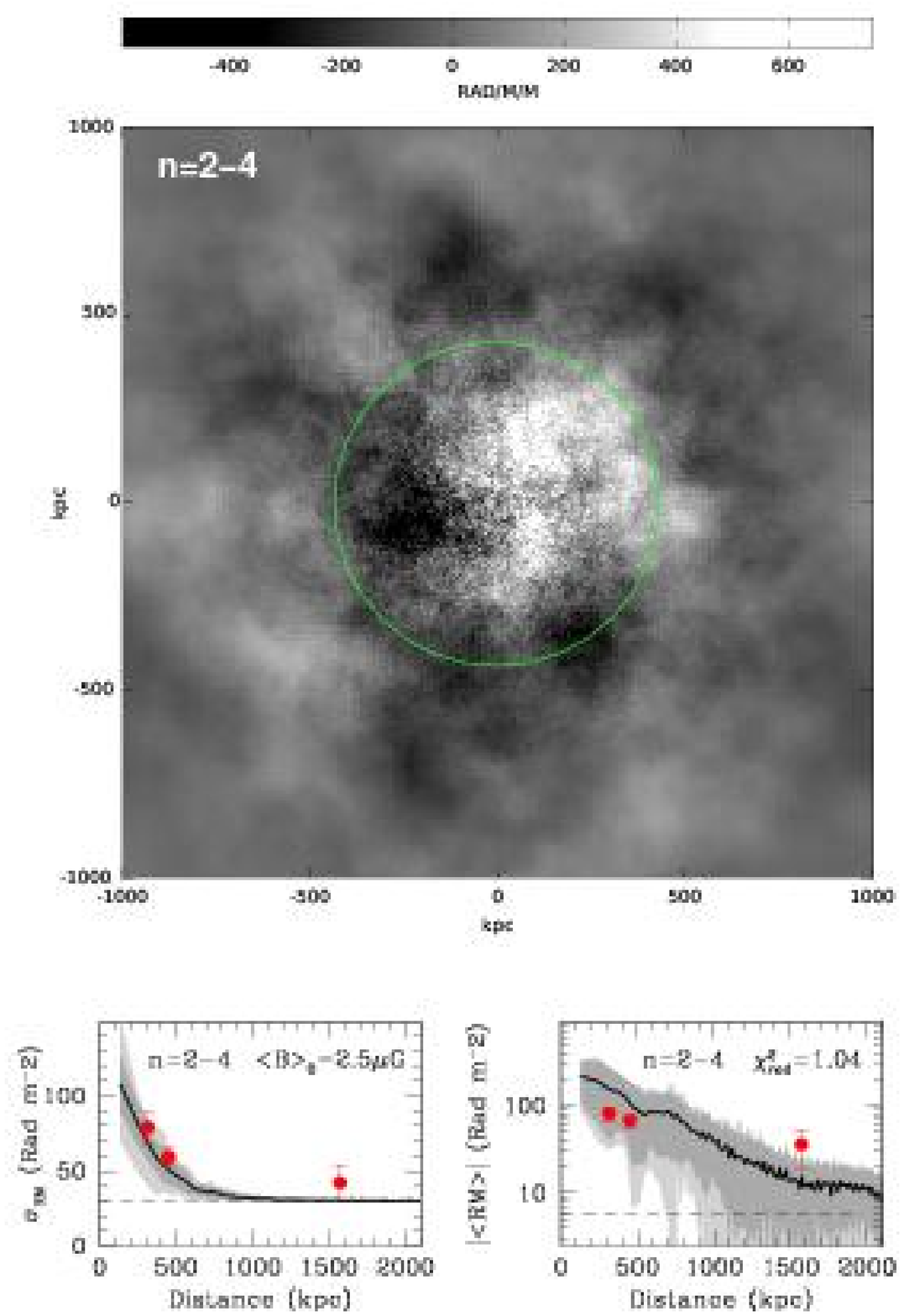}
\caption[]{
Top: simulated RM image as appears for a variable value of the magnetic 
field power spectrum spectral index $n=2-4$. 
In this simulation the average field at the cluster center 
is \bm$_{\rm0}$ $=2.5 ~\mu$G and its  
energy density decreases from the cluster center 
according to $B^{2}\propto n_{\rm e}(r)$.
The RM image shows a field of view of about $2\times 2$ Mpc. 
We performed the RM integration  
from the cluster center up to 4 Mpc along the line-of-sight.
The circle represent the cluster core radius.
Bottom:
Comparison of \srm~(left column panel)
and \absrmm~(right column panel) data of the three cluster 
radio galaxies (dots) with the simulated 
profiles (dark lines). The dark gray regions
show the dispersion of the profiles calculated in the simulations.
The light gray regions represent the increase of the scatter of simulated RM in the case in which
the Faraday depth of the simulated RM images is extracted randomly 
from the 3 dimensional distribution of galaxies in A2255.
The horizontal dashed lines represent the lower limits of \srm~ and \rmm~
caused by the presence of fitting errors and the RM galactic contribution. In the right panel
 the value of the reduced $\chi^{2}$ for two degree of freedom is also indicated. 
}
\label{fit_rm_n2-4}
\end{figure*}

\subsection{Simulated Rotation Measure images} 
The resulting cluster RM images and profiles, obtained with the same procedure described in 
Sect. 6.1, are shown in Fig.~\ref{fit_rm_n2-4}. 
The average magnetic field strength at the cluster center is 2.5\,$\mu G$.
The RM image shows that the small
 scale field fluctuations are stronger at the cluster center while the large scale magnetic field 
dominates at the periphery. The RM profiles shows that this model is able to reproduce both the 
 \srm~ and the \rmm~ of all three sources better than a constant magnetic field slope model can.

\begin{figure*}
\centering
\includegraphics[width=16cm]{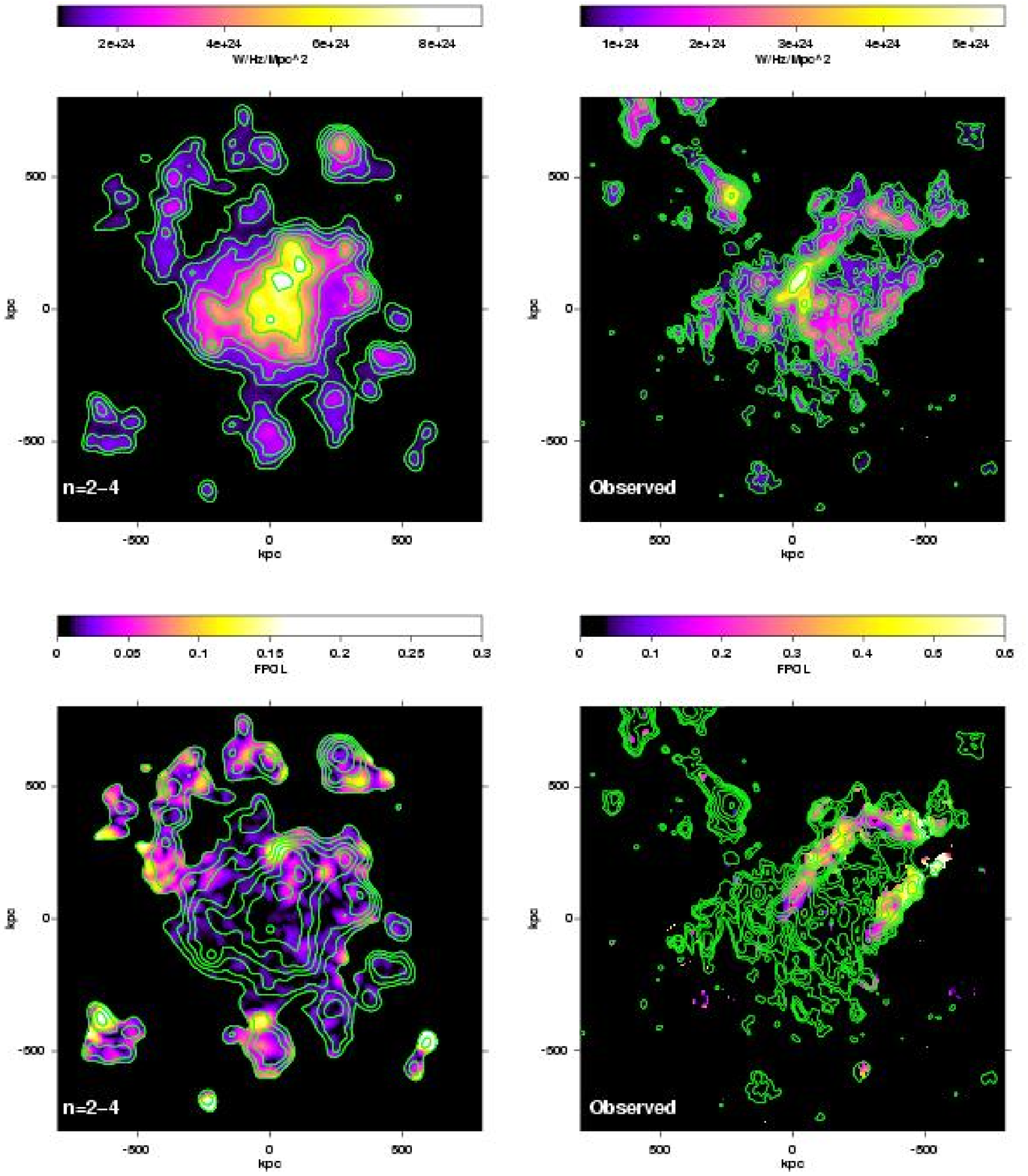}
\caption{
Top: 
Simulated (left panel) radio halo brightness for a variable magnetic field spectral index $n$=2-4. 
For a direct comparison with the data (right panel)
this image has been convolved with a beam of 25\arcsec~ 
(37.5 kpc) and the lowest contour level has been chosen to match 
the dynamic range of 10 as in the observations.
Bottom:
Simulated (left panel) radio halo brightness (contours) and fractional polarization (color) 
for the variable magnetic field spectral index $n$=2-4.
For a direct comparison with the data see the right panel.
Note that in the observed image of A2255 the radio galaxies have been subtracted and the emission 
of the peripheral radio relic is not in the field of view.
}
\label{halo_n2-4}
\end{figure*}

\subsection{Simulated radio halo images}

In Fig.~\ref{halo_n2-4}, we present the observed radio halo total intensity and
 fractional polarization images (right panels) along with their simulated counterparts (left panels) 
generated by considering the magnetic field model outlined above.
 By comparing the total intensity images we are able to
 reproduce both 
the brightness level,  
 the size, and the overall morphology of the halo.  More importantly, the fractional polarization of 
 the simulated filaments reaches values as high as 10\%-20\%. These values,
 although still somewhat lower than those observed, 
(the average polarization percentage in filaments is 30\%), 
are substantially higher than those presented in the previous section
 and clearly represent an improvement over the model based on a constant slope $n$.

Note that with the same magnetic field strength we are able to explain both the RM data and the 
radio luminosity of the radio halo while keeping relativistic particle and field energy densities 
in equipartition.
However, the equipartition magnetic field of the {\it simulated} radio halo 
calculated using standard formulas (Pacholczyk 1970) 
results in about 0.3\,$\mu G$ over a 1Mpc$^3$ volume
(the same result obviously holds also for the observed radio halo 
since both have roughly the same size and 
luminosity). With such a magnetic field strength, which is lower
 than the average magnetic field of the simulations over the same volume
 $\langle B\rangle \simeq 1.2$\, $\mu G$, it would be virtually impossible to fit the RM data, 
since the expected
 \srm~ and \rmm~ profiles will be one order of magnitude lower 
than those shown in Fig.\ref{fit_rm_n2-4}. 
This problem is usually addressed in the literature as a discrepancy between the magnetic field 
strength estimated from the RM and that estimated from the minimum energy argument.
In our simulation this problem is not present since most of the relativistic particles
 are at low energies and are undetectable at observable radio wavelengths.
We do not claim that we have solved the aforementioned discrepancy. But we note that 
the classical equipartition estimate is critically dependent on the low energy cut-off of the
electron energy spectrum for steep spectrum sources such as radio halos. 
Thus, until a precise knowledge of the low energy spectrum of the
synchrotron electrons 
in radio halos can be reached, the equipartition
 estimates of the magnetic field strength in these radio sources should be 
used with caution.

In conclusion, we find that a magnetic field power spectrum whose slope steepen from the cluster
 center outwards provides a good description of the RM radial profiles and is able to reproduce both the
luminosity and the fractional polarization of the radio halo, thus reconciling the magnetic field
 strength estimated obtained using these two complementary methods of analysis.

The fractional polarization of the observed radio halo is still higher than the simulated one. However,
 we do not exclude that more sophisticated magnetic field models, e.g. non-Gaussian random fields 
with ad-hoc correlated phases, can better reproduce the observed filamentary polarized structures.

We checked that the magnetic field radial profile adopted in our simulations,
in particular the value of $\mu$, does not strongly affect the results.

\section{Conclusions}
We present new VLA observations at 3.6 and 6\,cm 
for four polarized radio galaxies embedded in A2255, obtaining detailed RM
images for three of them. The dispersion and the mean of the RM decrease
with increasing distance from the cluster center.
We analyze these data, together with the very deep radio halo image recently 
obtained by Govoni et al. (2005). Using the numerical approach described in 
Murgia et al. (2004), we simulate random 3-dimensional magnetic 
field models characterized by different power spectra and produce synthetic 
RM and radio halo images. By comparing the simulations with the data we investigate 
the strength and the power spectrum of the intra-cluster magnetic field fluctuations. 

We find that a magnetic field power spectrum with a single spectral
index is not suitable to reproduce both the rotation measure
distribution of radio galaxies, and the radio halo polarization.
The data require a steepening of the power spectrum spectral index
from $n=2$, at the cluster center, up to $n=4$ at the cluster
periphery and the presence of filamentary structures on large scale.
This result seems to indicate that the magnetic field at
the cluster center would be more dominated by small-scale structures,
while at the periphery most of the magnetic field energy density is
concentrated on the large-scale structures.
A magnetic field with power spectrum steepening with radius
is likely to reflect a complex behaviour of turbulence and motions
in clusters.  
Subramanian et al. (2006) discuss the evolving
turbulence due to dynamo action, and find that turbulence can coexist
with ordered filamentary gas structure.
Since turbulence is likely driven by a merger event, it may be
expected that this driving is stronger in the cluster central
region. In this case, turbulence in the peripheral regions can be in a
more advanced stage of decay than at the cluster center.  Since
smaller scales decay first, and the integral scale of decaying
turbulence increases with time, larger-scale magnetic field structures
may be more easily present in the outer cluster region.

The average magnetic field strength at the cluster center is 2.5\,$\mu G$. 
The field strength declines from the cluster center outward and the average magnetic field 
strength calculated over 1 Mpc$^3$ is about 1.2\,$\mu$G. 

In conclusion, we find that the above magnetic field model provides a good description of the 
RM radial profiles and is able to reproduce both the luminosity and the fractional polarization
of the radio halo, thus reconciling the magnetic field
strength estimated using two complementary methods of analysis.

\begin{acknowledgements}
We are grateful to Dan Harris and Anvar Shukurov
for discussions and suggestions.
This research has made use of the
NASA/IPAC Extragalactic Data Base (NED) which is operated by the JPL, 
California Institute of Technology, under contract with the National 
Aeronautics and Space Administration.
This research was partially supported by PRIN-INAF 2005.
\end{acknowledgements}


\begin{thebibliography}{}
\bibitem{}
Burns, J.O., Roettiger, K., Pinkney, J., et al. 1995, ApJ 446, 583 
\bibitem{} 
Carilli, C.L. \& Taylor, G.B. 2002, ARA\&A, 40, 319
\bibitem{} 
Clarke, T.E., Kronberg, P.P., B{\" o}hringer, H., 2001, ApJ 547, L111 
\bibitem{} 
Clarke, T.E.: 2004, Journal of Korean Astronomical Society, 37, 337 
\bibitem{}
Condon J.J., Cotton W.D., Greisen E.W. et al., 1998, AJ 115, 1693
\bibitem{}
Dolag K., Schindler S., Govoni F., Feretti L., 2001, A\&A 378, 777
\bibitem{} 
Dolag, K., Bartelmann, M., Lesch, H., 2002, A\&A 387, 383 
\bibitem{} 
Dolag, K., Vogt, C., En{\ss}lin, T.A., 2005, MNRAS 358, 726 
\bibitem{}
En{\ss}lin T.A., Vogt C., 2003, A\&A, 401, 835 
\bibitem{}
Feretti, L., B\"ohringer, H., Giovannini, G., Neumann, D. 1997, A\&A 317, 432 
\bibitem{}
 Feretti, L., Dallacasa, D., Govoni, F., Giovannini, G., Taylor, G.B., Klein, U., 1999, A\&A 344, 472
\bibitem{}
Fusco-Femiano, R., Orlandini, M., Brunetti, G., Feretti, L., Giovannini, G., Grandi, P., Setti, G., 2004, ApJ 602, L73 
\bibitem{} 
Giovannini, G., Feretti, L., 2002, ASSL Vol.~272: Merging Processes 
in Galaxy Clusters, 197 
\bibitem{} 
Govoni, F., Falomo, R., Fasano, G., Scarpa, R.\ 2000, \aap, 353, 507 
\bibitem{}
 Govoni, F., Taylor, G.B., Dallacasa, D., Feretti, L., Giovannini, G., 
2001a, A\&A 379, 807
\bibitem{} 
Govoni, F., En{\ss}lin, T.A., Feretti, L., Giovannini, G., 2001b, 
A\&A, 369, 441 
\bibitem{} 
Govoni, F., Feretti, L., 2004, 
International Journal of Modern Physics D 13, 1549 
\bibitem{} 
Govoni, F., Murgia, M., Feretti, L., Giovannini, G., Dallacasa, D., Taylor, G.~B., 2005, A\&A 430, L5 
\bibitem{}
 Harris, D.E., Kapahi, V.K., Ekers, R.D., 1980, A\&AS 39, 215
\bibitem{}
 Jaffe, W.J., Rudnick, L., 1979, ApJ 233, 453
\bibitem{}
Miller N.A., Owen F.N., 2003, AJ 125, 2427
\bibitem{}
Murgia M., Govoni F., Feretti L., Giovannini G., Dallacasa D., Fanti R.,
Taylor G.B., Dolag K., 2004, A\&A 424, 429
\bibitem{}
Pacholczyk, A.G., Radio Astrophysics (Freeman, San Francisco, 1970)
\bibitem{}
Rephaeli, Y., Gruber, D., Arieli, Y., 2006, ApJ in press, astro-ph/0606097  
\bibitem{} 
Sakelliou, I., \& Ponman, T.~J., 2006, MNRAS 367, 1409 
\bibitem{}
Simard-Normandin M., Kronberg P.P., Button S., 1981, ApJS 45, 97
\bibitem{}
Struble, M.F., Rood, H.J., 1999, ApJS, 125, 35
\bibitem{}
Subramanian K.,  Shukurov A.,  Haugen  N.E.L., 2006, MNRAS 366, 1437
\bibitem{}
 Taylor G.B., Govoni F., Allen S., Fabian A.C., 2001, MNRAS 326, 2
\bibitem{}
Vikhlinin, A., Markevitch, M., 2002, Astronomy Letters, 28, 495 
\bibitem{} 
Yuan, Q., Zhou, X., \& Jiang, Z.\ 2003, ApJS, 149, 53 
\bibitem{} 
Yuan, Q., Zhao, L., Yang, Y.,  et al.\ 2005, AJ, 130, 2559 
\end{thebibliography}
\end{document}